\documentclass[useAMS,usenatbib]{mn2e}
\usepackage{graphicx}
\usepackage{natbib}
\usepackage{amssymb}
\usepackage{setspace}
\usepackage{amssymb}
\usepackage{epsfig,color}
\usepackage{multicol}

\long\def\symbolfootnote[#1]#2{\begingroup%
\def\thefootnote{\fnsymbol{footnote}}\footnote[#1]{#2}\endgroup} 

\newcommand{\Msol}{~$M_{\odot}$} % solar mass
\newcommand{\sqc}{cm$^{-2}$}                   % cm^-2
\newcommand{\cc}{cm$^{-3}$}                    % cm^-3

\newcommand{\thirthCO}{$^{13}$CO~}

\newcommand{\Hi}{\mbox{H\,{\sc i}~}}
\newcommand{\Hii}{\mbox{H\,{\sc ii}~}}

\title[General MC structure using {\sc Planck} data: I.]{Tracing the general structure of Galactic molecular clouds using {\sc Planck} data: I. The Perseus region as a test case}
\author[Stanchev et al.]
{
\parbox{\textwidth}{Orlin Stanchev$^1$\thanks{E-mail: \texttt{o\_stanchev@phys.uni-sofia.bg}}, Todor V. Veltchev$^{1,2}$, Jens Kauffmann$^3$, Sava Donkov$^4$, Rahul Shetty$^2$, Bastian K\"{o}rtgen$^5$ and Ralf S. Klessen$^2$}\vspace{0.4cm}\\
 \parbox{\textwidth}{
   $^1$University of Sofia, Faculty of Physics, 5 James Bourchier Blvd., 1164 Sofia, Bulgaria\\
   $^2$Universit\"at Heidelberg, Zentrum f\"ur Astronomie, Institut f\"ur Theoretische Astrophysik, Albert-Ueberle-Str. 2, 69120 Heidelberg, Germany\\
   $^3$Max-Planck-Institut f\"ur Radioastronomie, Auf dem H\"ugel 69, D-53121 Bonn, Germany\\
   $^4$Department of Applied Physics, Technical University, 8 Kliment Ohridski Blvd., 1000 Sofia, Bulgaria\\
   $^5$Universit\"at Hamburg, Hamburger Sternwarte, Gojenbergsweg 112, 21029 Hamburg, Germany}
}

\date{Accepted 2015 May 04. Received 2015 May 04; in original form 2013 December 13}

\pagerange{\pageref{firstpage}--\pageref{lastpage}} \pubyear{2015}

\begin{document}
\label{firstpage}
\maketitle

\begin{abstract}
We present an analysis of probability distribution functions (pdfs) of column density in different zones of the star-forming region Perseus and its diffuse environment based on the map of dust opacity at 353 GHz available from the {\sc Planck} archive. The pdf shape can be fitted by a combination of a lognormal function and an extended power-law tail at high densities, in zones centred at the molecular cloud Perseus. A linear combination of several lognormals fits very well the pdf in rings surrounding the cloud or in zones of its diffuse neighbourhood. The slope of the mean density scaling law $\langle \rho\rangle_L \propto L^\alpha$ is steep ($\alpha=-1.93$) in the former case and rather shallow ($\alpha=-0.77\pm0.11$) in the rings delineated around the cloud. We interpret these findings as signatures of two distinct physical regimes: i) a gravoturbulent one which is characterized by nearly linear scaling of mass and practical lack of velocity scaling; and ii) a predominantly turbulent one which is 
best described by steep velocity scaling and by invariant for compressible turbulence $\langle\rho\rangle_L u_L^3/L$, describing a scale-independent flux of the kinetic energy per unit volume through turbulent cascade. The gravoturbulent spatial domain can be identified with the molecular cloud Perseus while a relatively sharp transition to predominantly turbulent regime occurs in its vicinity.
\end{abstract}


\section{Introduction}
One of the tasks of the High Frequency Instrument \citep[HFI; see][]{Lamarre_ea_10} operating on the {\sc Planck} satellite is to map the spatial distribution of Galactic molecular and diffuse gas. The data on dust emission at frequency of 353 GHz allow for tracing the general structure of molecular clouds (MCs) and star-forming regions in a wide range of spatial scales -- from vast zones that include MCs and their neighbourhood at scales of tens of parsec through their diffuse envelopes down to their dense and clumpy central zones. Such a study would contribute to the numerous attempts of different authors to model and explain the MC structure through analysis of the column density distribution \citep{LAL_10, BP_ea_12}, of discrete, hierarchical clumps \citep{Kauffmann_ea_10a, Kauffmann_ea_10b} or combining both approaches \citep{Beaumont_ea_12}.

The Perseus region is particularly interesting for the study of general MC structure due to several reasons: i) it is compact, apparently axisymmetric; ii) it does not overlap with other fore- or background clouds or cloudlets, due to its high galactic latitude ($b\sim-20^\circ$); iii) it is associated with several sites of recent or active star-formation; iv) the whole region has been extensively investigated \citep[see][for a review]{Bally_ea_08}. 

\begin{figure*} 
\begin{center}
\includegraphics[width=1.\textwidth]{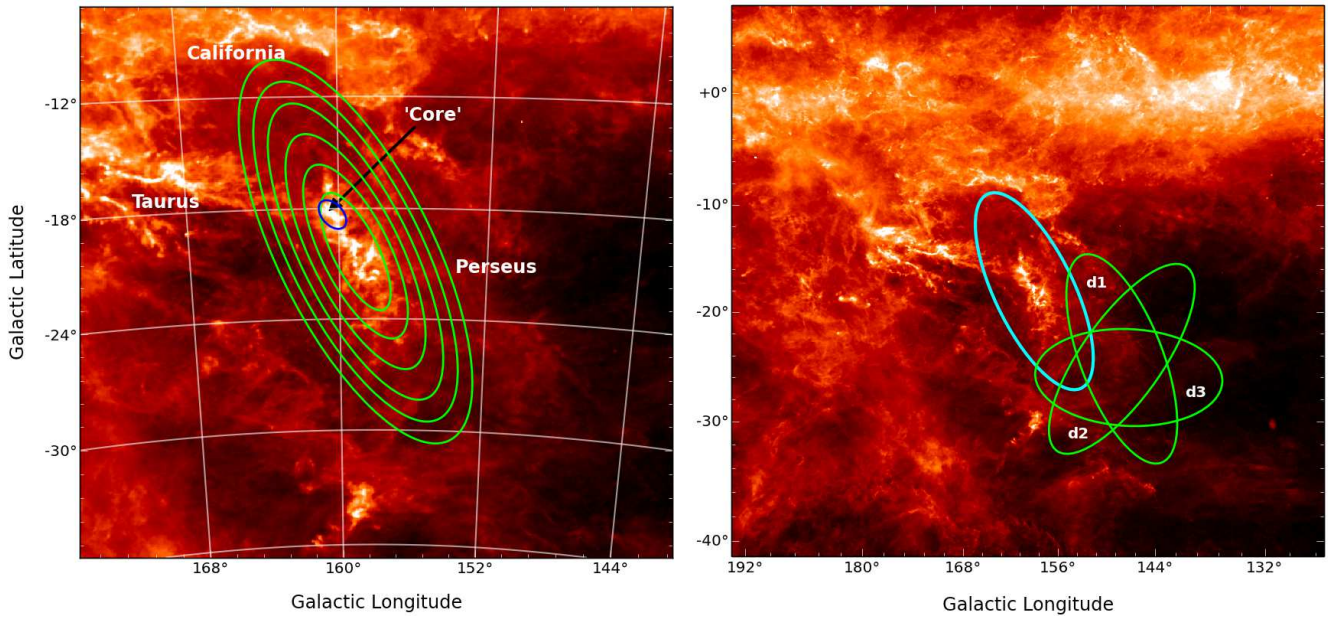}
\vspace{0.2cm}  
\caption{The selected probe zones, drawn on the {\sc Planck} dust opacity map: {\it (left)} the `Perseus region' (PR), containing Perseus MC (the central ellipse) and a high-extinction `core'; {\it (right)} zones of diffuse gas (labelled; see also Table \ref{table_ezones_param}) in the neighbourhood of the PR (unlabelled)}.
\label{fig_probe_regions}
\end{center}
\end{figure*}
The shape of the column density probability distribution function (hereafter, $N$-pdf) contains important information about the physics of star-forming regions. Dust extinction \citep{Kainulainen_ea_09, LAL_11} as well dust continuum studies \citep[e.g.][]{Schneider_ea_13} demonstrate that the $N$-pdf in Galactic MCs is nearly lognormal in some cases or can be fitted through a combination of lognormal and power-law functions, in other cases. A purely lognormal $N$-pdf is typical for turbulent medium \citep[e.g.][]{VS_94} as the standard deviation depends on the Mach number and the forcing of turbulence \citep{FKS_08}. On the other hand, numerical simulations of self-gravitating medium indicate that the development of a power-law (PL) tail is to be expected at high densities and at timescales, comparable to the free-fall time \citep{Klessen_00, KNW_11, FK_13, Girichidis_ea_14}. Therefore, the $N$-pdf can be used as an effective research tool to disentangle complex issues like scaling laws in turbulent MCs, 
energy equipartitions in star-forming regions and their relation to cloud evolutionary status. 

In this Paper we study and compare the behaviour of the $N$-pdf in embedded zones of Perseus region, containing Perseus MC, and in zones of its diffuse vicinity. The performed analysis allows for the reconstruction of the density scaling law and for the description of the physical regime in the Perseus region.

\section{Observational data and probe zones}
We used the component map of dust opacity at frequency 353 GHz (angular resolution:  $1.\!^{\prime}72$ per pixel) available from the {\sc Planck} archive\footnote{http://irsa.ipac.caltech.edu/data/Planck/release\_1/all-sky-maps/previews/HFI\_CompMap\_DustOpacity\_2048\_R1.10} to extract a large area containing the Perseus MC. Dust emission data are more appropriate for our study than dust extinction maps for two major reasons. First and foremost, extinction mapping depends on detection of background stars and thus suffers from selection effects. Precise photometry of weak stellar sources is possible in zones of low to moderate extinction which biases extinction maps towards lower values. Consequently, masses of fragments of fixed size can be essentially underestimated \citep[cf. Sec. 2.2.2 in][]{Kauffmann_ea_10b}. Second, {\sc Planck} data offer the opportunity to probe regions of very high column densities, not resolved in dust extinction maps due to a loss of background stars. 

\begin{table}
\caption{Statistics of the selected elliptical zones.}
\label{table_ezones_param} 
\begin{center}
\begin{tabular}{cccc@{~}c}
\hline 
\hline 
Zone & Nr. & R & $\langle N\rangle_R$ & Mass \\ 
~ & of pixels  & [ pc ] & [ $10^{21}$\sqc~] & [~$10^4$\Msol~] \\ 
\hline 
\multicolumn{5}{l} {\it Embedded zones in the `Perseus region'} \vspace{4pt}\\
`Core' & ~~4663 & ~5.3	& 10.1	& ~3.4 \\	
1 (the MC) & ~14498 & ~9.1	& ~7.4	& ~7.8 \\	
2 & ~32299 & 13.6	& ~5.2	& 12.0 \\	
3 & ~57339 & 18.1	& ~3.9	& 16.0 \\	  
4 & ~89649 & 22.6	& ~3.1	& 20.0 \\	
5 & 123812 & 26.5	& ~2.8	& 25.0 \\	
6 (the PR) & 163351 & 30.4	& ~2.5	& 30.0\vspace{6pt}\\	
\multicolumn{5}{l} {\it Diffuse zones in the neighborhood} \vspace{4pt}\\
d1 & 155132 & 30.4	& ~1.0	& 14.7 \\	 
d2 & 155121 & 30.4	& ~0.8	& 10.0 \\	 	
d3 & 147628 & 30.4	& ~0.9	& 10.0 \\	
d1 $\cup$ d2 $\cup$ d3 & 318479 & 41.4	& ~0.9	& 19.6 \\	
\hline 
\hline 
\end{tabular} 
\end{center}
\smallskip 
\end{table}

\begin{figure*} 
\begin{center}
\includegraphics[width=1.\textwidth]{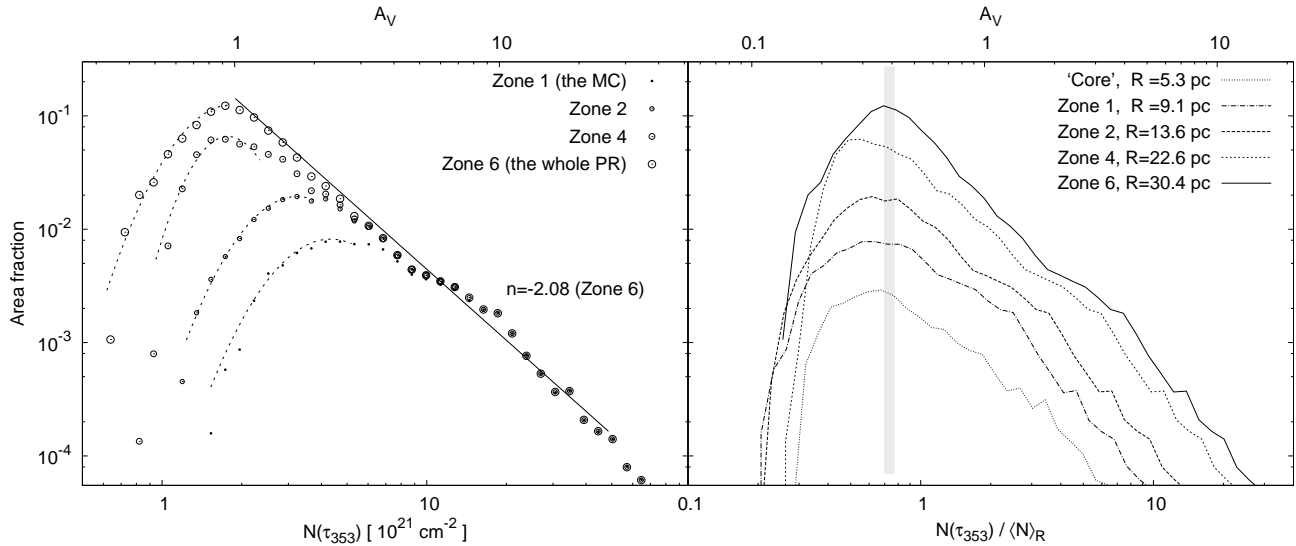}
\vspace{0.3cm}  
\caption{Column density pdfs in the PR (Fig. 1): {\it (left)} $N$ is in absolute units, the lognormal fits of all pdfs (dashed) and the fit of the PL tail in the largest zone (solid) with its slope $n$ are given; {\it (right)} $N$ is normalized to its mean value in the considered zone; the span of $N_{\rm low}^{\rm PL}$ is plotted with vertical shaded area. Only some of the pdfs are shown for clarity.}
\label{fig_N-pdf_cloud}
\end{center}
\end{figure*}

Recent research indicates that a distance gradient exists across the Perseus region \citep{Schlafly_ea_14}. Therefore all spatial sizes used in this Paper were calculated adopting a mean distance $D=260\pm40$~pc to the Perseus MC taking into account the parallax estimates to its western part \citep{Hirota_ea_11} and the variety of other ones to different parts of it \citep[for review, see][]{Bally_ea_08}. 

The selected zones to derive the $N$-pdf are shown in Fig. \ref{fig_probe_regions} while general physical information on them is provided in Table \ref{table_ezones_param}. The zone labelled `Perseus region' (hereafter, PR) contains six embedded ellipses of increasing size and identical orientation angle, following the morphology of Perseus MC (Fig. \ref{fig_probe_regions}, left). The central ellipse (\#1) encompasses the very cloud. The next two (\#2 and \#3) cover the cloud and also filamentary structures connected to it. The three largest elliptical zones (\#4, \#5 and \#6) include the cloud and its filamentary vicinity plus increasing areas of diffuse gas and of the south-eastern part of the Perseus OB2 association, but without overlapping with the California MC (outside the northern edge of Zone \#6). The dense molecular gas with $N\gtrsim10^{22}$~\sqc~ is concentrated in Zone \#1 (the Perseus MC) wherein we delineated also a `core' with high dust extinction and associated with the \Hii region G159.6-18.5 \citep{Bally_ea_08}. To compare the contribution of diffuse gas (atomic and molecular) to the $N$-pdf, we selected additionally three zones in the western neighbourhood of PR with the same areas like the largest elliptical zone (\#6, i.e. the whole PR) and spanning similar column density ranges (Fig. \ref{fig_probe_regions}, right; Table \ref{table_ezones_param}, bottom). 

A standard way to estimate column density from observations of interstellar medium is from extinction measurements, adopting a constant ratio of column density $N$ to visual extinction $A_V$. The value of $N(\rm H\,${\sc i}$)/A_V$ which is frequently used in different works is $1.9\times10^{21}$~\sqc~ \citep{BSD_78}. It has been obtained from photometry of OB stars, with an upper limit in the chosen sample $E(B-V)\simeq0.\!\!^m8$~or, $A_V \lesssim 2.\!\!^m5$. In view of this limitation, transformations to column density in MC fragments of high extinction (typically populating the PL tail of the $N$-pdf) should be considered with caution. Therefore, in this Paper, we derive $N{(\rm H)}$ directly from dust opacity $\tau_{353}$ at 353 GHz, provided from {\sc Planck} data. As evident from Fig. 6 in \citet{Planck_11}, there is a very good linear correlation between dust opacity and total hydrogen column density $N$(H) obtained from \Hi 21~cm maps, except for the range $0.7\times 10^{21}\le N({\rm H})\le 5 \times 
10^{21}$\sqc~wherein undetected (`dark') molecular gas may affect the results. Neglecting the latter effect, we adopted a linear conversion formula from dust opacity to hydrogen column density:
\begin{equation}
 \label{eq_tau_to_N}
 N({\rm H})=C_1 \tau_{353} + C_0~,
\end{equation}
where coefficients $C_0$ and $C_1$ were obtained from fitting the $N({\rm H})$-binned representation of the correlation at column densities below $0.7\times 10^{21}$\sqc~(see Fig. 6 in Planck Collaboration 19, 2011). The possible uncertainty of the calculated $N({\rm H})$ is about a factor of 2 which is not significant for the analysis presented in this Paper.

\section{Column density pdfs}
\subsection{Perseus region}
The derived $N$-pdfs in the PR and in its embedded zones (Fig. \ref{fig_probe_regions}, left) are plotted in Fig. \ref{fig_N-pdf_cloud} (left). Apparently, their shape can be described by a combination of a lognormal function around the peak and a power-law (PL) tail with slope varying in a narrow range between $-1.7$ and $-2.1$. Such behaviour is in agreement with other observations of nearby MC complexes \citep{Kainulainen_ea_09, FK_13, Schneider_ea_14a, Schneider_ea_14b} or with results from numerical simulations of self-gravitating clouds \citep{KNW_11}. The slope and the lower column-density limit $N_{\rm low}^{\rm PL}$ of the PL tail were calculated through the method {\sc Plfit} \citep{CSN_09} which extracts these characteristics from analysis of the {\it unbinned} observational data. In this way we avoid subjective estimations -- note, that the obtained values $N_{\rm low}^{\rm PL}$ (Table \ref{table_pdf_param}) are lower than one would put `by eye' (say, at $\sim8\times10^{21}$~\sqc; Fig. \ref{fig_N-pdf_cloud}, left). 

Normalisation to the mean column density $\langle N \rangle$ in each considered zone yields $N$-pdfs with practically identical shape and $N_{\rm low}^{\rm PL}/\langle N\rangle$ (Fig. \ref{fig_N-pdf_cloud}, right), for a wide range of effective radii. Physical interpretation of this result is suggested in Sect. \ref{Physical_analysis}.

\begin{figure*} 
\begin{center}
\includegraphics[width=.85\textwidth]{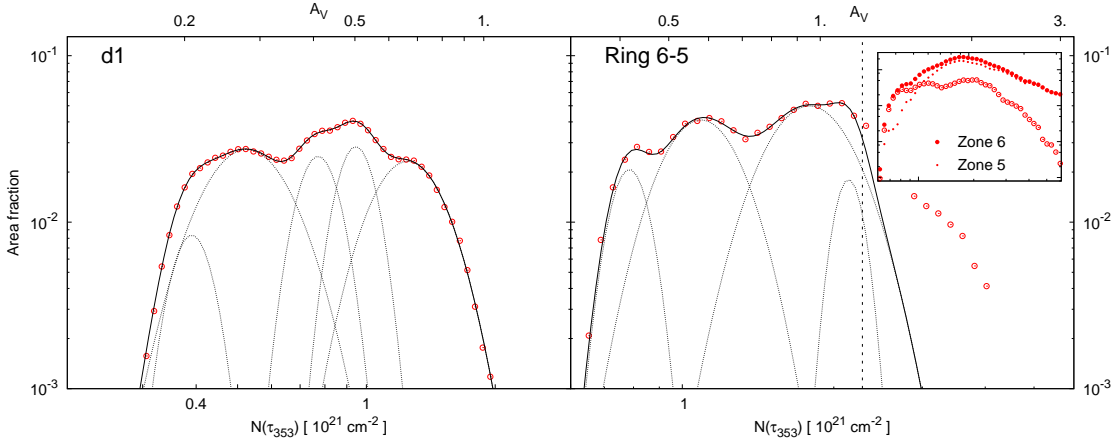}
\vspace{0.3cm}  
\caption{Examples of $N$-pdf (open symbols) decomposition to lognormals (dotted) in a diffuse zone (left) and in a ring (right) delineated by the boundaries of two successive elliptical zones in the PR (cf. Table \ref{table_ezones_param}). In the latter case, the location of the lower boundary $N_{\rm low}^{\rm PL}$ of the PL tail is indicated (dashed). The embedded graph demonstrates how the $N$-pdf in a considered ring results from $N$-pdfs of the elliptical zones (filled symbols).}
\label{fig_pdfs_decomposition_examples}

\end{center}
\end{figure*}

\subsection{Diffuse zones}
\label{PDF_analysis_diffuse_zones}
The $N$-pdf in each diffuse zone exhibits a complex shape with several discernible peaks in whose vicinity the distribution is apparently lognormal (Fig. \ref{fig_pdfs_decomposition_examples}, left). The same behaviour is found when one considers the $N$-pdf in any of the rings outlined by the boundaries of embedded elliptical zones in the PR, excluding the PL tail (Fig. \ref{fig_pdfs_decomposition_examples}, right).  

\subsubsection{Method for pdf decomposition}
\label{Fitting_procedure}
The column-density distribution in any of these regions of diffuse gas can be represented as sum of lognormal functions of type

\[ {\rm lgn}_i(N; a_i, N_i, \sigma_i)=\frac{a_i}{\sqrt{2 \pi\sigma_i^2}}\exp\!\Bigg(\!\!-\!\frac{[\,\lg(N/N_i)\,]^2}{2 \sigma_i^2}\!\Bigg), \]

\noindent where the total number $m$ of components ($1 \le i\le m$) spans typically from a few to dozen and the parameters $N_i$, $\sigma_i$ and $a_i$ are obtained from the following fitting procedure of the whole distribution: 
\begin{itemize}
 \item {\it Prominent local peaks}, i.e. of considerable width and height, are located and guess values $N_i^{(0)}$ are assigned to them. It is appropriate to start from such peaks with broad wings (large $\sigma_i$) since they have greater statistical weight.
 \item The distribution is fitted {\it in close vicinities of these peaks} through single lognormals with fitting parameters $N_i^{(1)}$, $\sigma_i^{(1)}$ and $a_i^{(1)}$.
 \item The total fitting function $\sum_{i=1}^m {\rm lgn}_i$ is composed. Then the {\it whole} distribution is fitted and next approximations of the fitting parameters $N_i^{(2)}$, $\sigma_i^{(2)}$ and $a_i^{(2)}$ are obtained.
 \item {\it Small local peaks and inflexion points} are located and the corresponding number of lognormal components ${\rm lgn}_i$ is added to the fitting function, increasing $m$. The previous three steps are repeated and a next set $N_i^{(3)}$, $\sigma_i^{(3)}$ and $a_i^{(3)}$ is calculated for {\it all} components. 
 \item In case the achieved approximation of the distribution is not satisfactory, a few further components could be added. The latter have small statistical weights and contribute only for {\it small local improvements} of the fit. 
 \item Now the fitting-parameters' space $(N_i, \sigma_i, a_i)$ $(i=1,...,m)$ is severely restricted and the total fitting function $\sum_{i=1}^m {\rm lgn}_i$ is fitted via NLLS Levenberg-Marquardt algorithm until the $\chi^2$ criterion for fit's goodness (at 95\% confidence level) is satisfied. 
\end{itemize}

All obtained fits in the diffuse zones and in the rings of the PR are plotted in the Appendix \ref{Appendix_decomposition}, Figs. \ref{fig_N-pdf_dif} and \ref{fig_N-pdf_rings}. The achieved fits' goodness is given in Table \ref{table_fit_goodness_obs}. 

\subsubsection{Lognormal components and spatial scales}
\label{Components_and_scales}
\begin{figure} 
\begin{center}
\includegraphics[width=85mm]{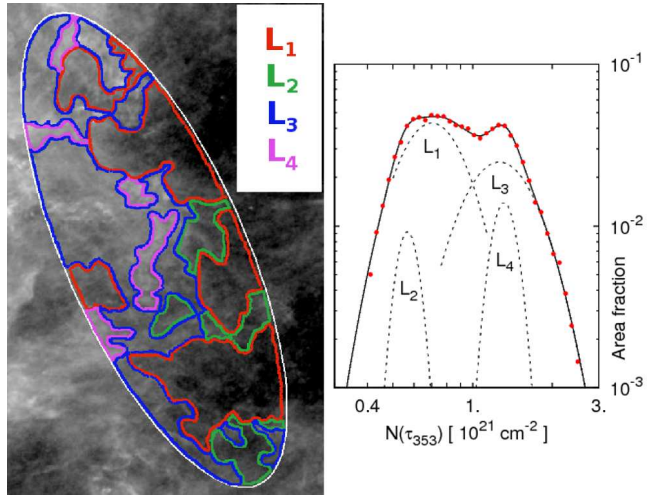}
\vspace{0.2cm}  
\caption{Schematic representation of spatial scales $L_i$ in our approach. The border of a considered diffuse region is drawn (white line). The positions and shapes of the fragments are arbitrary and represent one possible spatial realization of the scales.}
\label{fig_scheme_scales}
\end{center}
\end{figure}

A possible interpretation of this decomposition of the $N$-PDF in a region to lognormals is that domains of various typical column densities $N_i$ and velocity dispersions (related to $\sigma_i$; see e.g. Sect. 3.6 in \citealt{Federrath_ea_10}) contribute to it. Each domain is constituted of non-overlapping fragments of the considered region. Through the third parameter $a_i$ of a given component, we may define the effective size $L_i$ of a domain and hence introduce the notion of {\it spatial scale}. Each scale has its own lognormal column-density pdf ${\rm lgn}_i(N)$ and fractional area $\int {\rm lgn}_i(N)/\sum_i \int {\rm lgn}_i(N)$. Then the effective sizes of scales included in a zone of diffuse 
gas with total area $S$ and effective radius $R=\sqrt{S/\pi}$ are derived by:
\begin{equation}
 \label{eq_eff_size_scale}
  L_i=\sqrt{\frac{a_i}{\sum\limits_i a_i}}\,R~.
\end{equation}
We stress that spatial scale is a rather abstract quantity which includes, in general, a {\it set of fragments} with various sizes (down to single pixels) and shapes. The notion is illustrated in Fig. \ref{fig_scheme_scales} which represents one possible spatial realization of four scales ``detected'' through decomposition of the $N$-pdf in an exemplary diffuse region. The larger scales $L_1$ and $L_3$ (large $a_i/\sum_i a_i$) consist of fragments spanning broad column density ranges (wide $N$-pdfs) and have different typical densities $N_i$. On the other hand, the smaller scales $L_2$ and $L_4$ (smaller $a_i/\sum_i a_i$) are more homogeneous (narrow $N$-pdfs), with minor variations of $N$. 

The $N$-pdfs in the PR cannot be represented as combinations of lognormals but note that their column density ranges almost entirely fall into the one, corresponding to the PL tail of the largest zone \#6 (Fig. \ref{fig_N-pdf_cloud}, left). Therefore one may consider zones \#1 to \#5 as subscales of \#6, adopting their effective radii $R$ as spatial scales. It will be shown below that all zones in the PR (except the cloud `core') obey a mass scaling law with identical slope. 

\section{Physical analysis}
\label{Physical_analysis}
\subsection{Assumptions and definitions}
\label{Assumptions}
The analysis presented below is based on several reasonable physical assumptions. Some of them are discussed and justified in \citet{DVK_11}, in the framework of a statistical approach for description of MC structure at an early evolutionary stage. Here we briefly summarise them as follows: 

\noindent(i) Power scaling laws of velocity dispersion and mean density \citep[often called ``first and second Larson's laws'';][]{Larson_81}, as well of mass:
 \begin{equation}
 \label{eq_scaling_law_velocity}
  u_L\propto L^\beta~,~~~\beta>0~.
 \end{equation}
 \begin{equation}
 \label{eq_scaling_law_density}
 \langle\rho\rangle_L \propto L^\alpha~,
 \end{equation}
 \begin{equation}
 \label{eq_scaling_law_mass}
 M_L\propto L^\gamma~.
 \end{equation}
 The natural relation $M_L=\langle\rho\rangle_L L^3$ leads to: 
  \begin{equation}
  \label{eq_gamma_alpha}
  \gamma=\alpha+3 .   
  \end{equation}

\noindent(ii) An equipartition relation between gravitational and turbulent energy of type $E_{\rm kin} \sim f|W|$, where $f$ is a constant\footnote{~Note that the value of $f$ is not important for this analysis, e.g. we do not assume gravitational boundedness ($f\le0.5$) or virialization of the considered volume.}. The existence of such energy equipartition at different spatial scales and evolutionary stages in a MC is observationally verified and is explained by the coupling of various physical processes in the interstellar medium to each other \citep[see Sect. 1 in][]{HF_12}. It is supported as well by numerical simulations \citep{VS_ea_07}.

Hence one obtains a relation between the scaling indices of mass and velocity dispersion:
 \begin{eqnarray}
  \label{eq_beta_gamma}
 \frac{1}{2}M_L u^{2}_L \propto \frac{3}{5}G \frac{M_L^{2}}{L}~\,\Rightarrow & ~ & ~ \nonumber \\
 u_L\propto \Big(\frac{M_L}{L}\Big)^{\frac{1}{2}}\propto L^{\frac{\gamma-1}{2}}~,&~~{\rm i.e.}~~~\beta=(\gamma-1)/2~&~ 
  \end{eqnarray}

How can one estimate the mean density at given spatial scale? The most straightforward approach is to define $\langle\rho\rangle_L$ as a quotient of the mean column density and the scale. At a scale $L_i$ in the diffuse rings and zones, characterized by a lognormal $N$-pdf with peak $N_i$, $\langle N \rangle$ is simply $N_i$. On the other hand, $N$-pdfs in zones of the PR are asymmetric and in this case we use the estimates of $\langle N \rangle$ from Table \ref{table_ezones_param} (col. 4) to calculate $\langle\rho\rangle_R\approx \langle N \rangle/R$. 

The mean-density scaling laws in the diffuse rings, in the elliptical zones and in the test simulation (see Sect. \ref{Comparison with simulations}) are derived from the $N$-pdfs via a weighted least-squares fitting described in Appendix \ref{Appendix_fitting}. These scaling laws and those of velocity dispersion hold in the inertial range of a supersonic turbulent cascade \citep{DVK_11}. Therefore it is instructive to check the behaviour of the flux of kinetic energy per unit volume through the cascade in case of compressible turbulence, defined as:

\begin{equation}
 \label{eq_kinetic_energy_flux}
 \langle \rho\rangle_L\frac{u_L^2}{\tau_L}=\langle \rho\rangle_L\,\frac{u_L^2}{L/u_L}=\langle \rho\rangle_L\,\frac{u_L^3}{L}
\end{equation}
where $\tau_L$ is the turbulent crossing time at scale $L$. The possible scaling of this quantity is part of the physical analysis presented below.

\subsection{Physical regimes in the Perseus region and its vicinity}
We suggest that the $N$-pdfs as well the underlying $\rho$-pdfs reflect two types of physical regime which govern the general structure of MCs. The PL tail corresponds to a regime in which gravity determines structure whereas the pdf part that can be decomposed to lognormal components corresponds to a regime with a prevailing role of turbulence. This concept is consistent with the fact that a well-developed PL tail, spanning about one order of magnitude or more, is present in pdfs of regions containing the very MC (Fig. \ref{fig_N-pdf_cloud}) while a pdf with lognormal components and without PL tail is characteristic of the rings in the PR and of the diffuse zones in its vicinity (Figs. \ref{fig_N-pdf_dif} and \ref{fig_N-pdf_rings}). The derived mean density scaling laws in the suggested two physical regimes also differ significantly -- they are illustrated in Fig. \ref{fig_n_scaling_cloud}. Let us examine the emerging picture closely.

\subsubsection{Zones forming the power-law tail of the $N$-pdf}
\label{PL_tail_zones}
As already commented in Sect. \ref{PDF_analysis_diffuse_zones}, the embedded zones in the PR can be considered as subscales of the largest Zone \#6 and thus the latter should determine the balance of energies in the cloud structure. The $N$-pdfs in these zones, normalized to the corresponding mean column densities $\langle N \rangle_R$, turn out to be strikingly identical (Fig. \ref{fig_N-pdf_cloud}, right, and Table \ref{table_pdf_param}), with the same width $\sigma_N$, peak $N_{\rm max}/\langle N \rangle_R$ and lower boundary $N_{\rm low}^{\rm PL}/\langle N \rangle_R$ of the PL tail. (Only the MC `Core' exhibits an exception from this trend.) This result might be interpreted by assuming that the Perseus MC is self-gravitating and its general structure is described by power-law surface- and volume-density profiles. 

\begin{table}
\caption{Parameters of the $N$-pdfs derived in the embedded elliptical zones of the PR. Column densities are given in units [$10^{21}$~\sqc]. Notation: $N_{\rm max}$ - peak of the distribution, $n$ - slope of the PL tail, $f_{\rm PL}(S)$ - PL area fraction, $f_{\rm PL}(M)$ - PL mass fraction.}
\label{table_pdf_param} 
\begin{center}
\begin{tabular}{ccccccc}
\hline 
\hline 
Zone & $N_{\rm max}$ & $\sigma_N$ & $N_{\rm low}^{\rm PL}$ & $n$ & $f_{\rm PL}(S)$ & $f_{\rm PL}(M)$\\ 
\hline 
`Core'	& 5.97	& 0.14 & 5.95	&  $-1.67$	&	 0.88	&	0.84	\\	
1	& 4.51	& 0.18 & 5.22	&  $-1.84$	&	 0.85	&	0.77	\\	
2	& 3.35	& 0.17 & 3.90	&  $-1.83$	&	 0.83	&	0.72	\\	
3	& 2.15	& 0.12 & 2.85	&  $-1.85$	&	 0.82	&	0.71	\\	
4	& 1.81	& 0.12 & 2.39	&  $-1.82$	&	 0.83	&	0.71	\\	  
5	& 1.80	& 0.13 & 1.98	&  $-1.93$	&	 0.84	&	0.73	\\	
6	& 1.89	& 0.16 & 1.85	&  $-2.09$	&	 0.85	&	0.74	\\	
\hline 
\hline 
\end{tabular} 
\end{center}
\smallskip 
\end{table}

Let a region with area $S$ and volume $V$, centred at $L=0$, is characterised by a surface-density profile $\Sigma(L)=\Sigma_{0}(L/1~{\rm pc})^{-t}$ and a volume-density profile $\rho(L)=\rho_{0}(L/1~{\rm pc})^{-p}$. This yields scaling laws of column density $\langle N\rangle\propto L^{\alpha_N}$ ($\alpha_N=-t$) and mean density $\langle\rho\rangle_L\propto L^{\alpha}$ ($\alpha=-p$), where those quantities are averaged over the entire profile. Then, if the column- and volume density pdfs in the region display PL tails, one gets for their slopes $n$ and $q$, respectively: 
\begin{equation}
 \label{eq_Sigma-N_PDF}
  p_N(\ln(\Sigma/\Sigma_0))\equiv \frac{dS}{d\ln(\Sigma/\Sigma_0)}\propto\Sigma^{-2/t}~~\Longrightarrow~~n=-2/t
\end{equation}
\begin{equation}
 \label{eq_rho-rho_PDF}
 p(\ln(\rho/\rho_0))\equiv \frac{dV}{d\ln(\rho/\rho_0)}\propto\rho^{-3/p}~~\Longrightarrow~~q=-3/p
\end{equation}

As demonstrated by \citet{KNW_11}, the relation between the volume-density and the surface-density profiles in isotropic medium is:
\begin{eqnarray}
 \Sigma(L) & = & \int\limits^{+\infty}_{-\infty}\,\rho((L^{2}+x^{2})^{1/2})\,dx = ... \nonumber\\
  ~ & \propto & L^{1-p}\,\int\limits^{+\infty}_{0}\,(1+(x/L)^{2})^{-p/2}d(x/L)~, 
\end{eqnarray}
 where the integral is converging for $1 \leq p \leq \infty$. Obviously, $t=p-1=-(\alpha+1)$. 

\begin{figure} 
\begin{center}
\includegraphics[width=83mm]{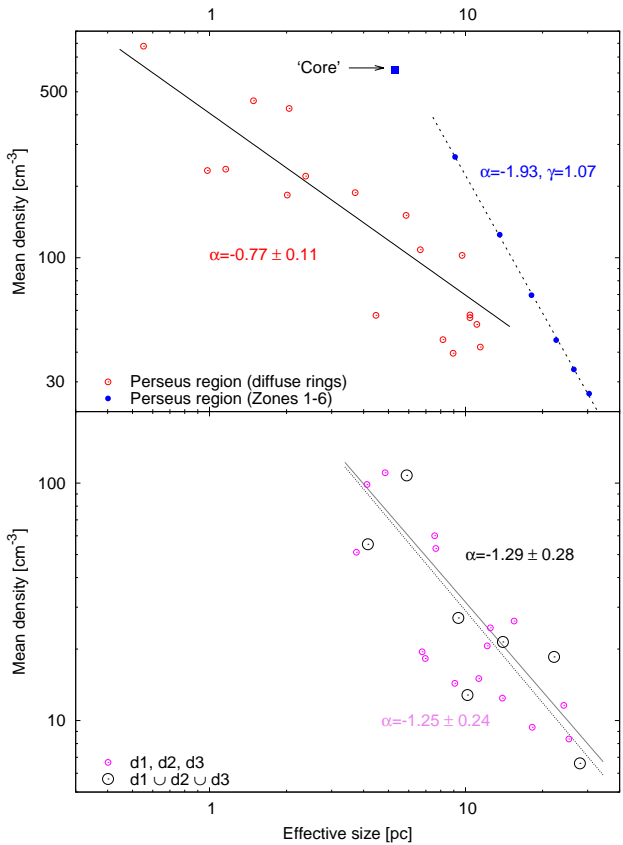}
\vspace{0.2cm}  
\caption{Power-law scaling of mean density: {\it(Bottom)} in the diffuse zones outside the PR (small circles) and in their union (large circles); {\it(Top)} in the PR (filled circles) and in its diffuse rings (open circles). The slope in the former case was derived excluding the cloud `core' (square).}
\label{fig_n_scaling_cloud}
\end{center}
\end{figure}

Now the derived mean density scaling index $\alpha=-1.93$ in the PR (Fig. \ref{fig_n_scaling_cloud}, top; blue symbols) leads through equation \ref{eq_Sigma-N_PDF} to a PL-tail slope $n=-2.15$, in a very good agreement with the value $-2.08$ obtained directly from the observational $N$-pdf (Fig.~\ref{fig_N-pdf_cloud}, left). On the other hand, one gets from equation \ref{eq_gamma_alpha} a mass-size scaling index $\gamma=1.07$.  Recalling assumption (ii) for equipartition of energies, we derive a velocity scaling law of vanishing slope (cf. equation \ref{eq_beta_gamma}): 
\begin{equation}
 \label{eq_no_scaling_velocity}
 u_L\propto L^{(\gamma-1)/2} \propto L^\beta~,~~~\beta\approx0.04
\end{equation}

This surprising result is consistent with the dendrogram analysis of cloud fragments from numerical simulations of self-gravitating cloud \citep[][see Fig. 2 in \citealp{Stanchev_ea_13}]{Shetty_ea_10}. Its physical meaning is that the sum of the specific kinetic and gravitational energies at each scale within the density span of the pdf tail is an invariant:
\begin{equation}
 \label{eq_invariant_MC}
 \frac{1}{2}u^{2}_{L}-\frac{3}{5}G\frac{M_L}{L}={\rm inv}(L)~.
\end{equation}
Note that this relation is {\it not} equivalent to the assumed equipartition of energies at each scale, although derived from it. The kinetic and gravitational energy terms contain as well the mass which is scale-dependent.

From equation \ref{eq_invariant_MC} one obtains for the flux of kinetic energy $\langle \rho\rangle_L u_L^3/L\propto L^{-3}$. This finding means that the kinetic flux increases strongly at small scales which can be naturally explained with matter acceleration due to gravity. Our result is very different from the one of \citet{GB_11} who derived from theoretical considerations of compressible isothermal turbulence $\langle\rho\rangle_L u_L^3/L={\rm inv}(L)$ for purely solenoidal forcing and $\langle\rho\rangle_L u_L^3/L\propto L^{2/3}$ for strong compressive forcing. The latter relation was confirmed as well by \citet{Federrath_13} from a number of high-resolution numerical simulations of compressible turbulence without gravity. It is evident that the structure of the PR centred on Perseus MC cannot be modelled taking into account only interstellar turbulence; self-gravity is an essential factor in constructing the correct physical picture.

\subsubsection{Diffuse zones within the Perseus region}
\label{Rings_in_Perseus_region}
The derived mean density scaling law in the rings of the PR is shown in Fig. \ref{fig_n_scaling_cloud}, top (red symbols). It is shallower than the classical ``second Larson's law'' with $\alpha=-1.1$ for MCs and cloud fragments \citep{Larson_81}. Taking the density scaling index $\alpha\simeq-0.80$, one gets $\gamma=2.20$ and $\beta\approx0.60$. Then -- in comparison with the analysis in the previous Section, -- we find here a scale-invariant flux of the total energy per unit volume:
\begin{equation}
 \label{eq_invariant_rings}
 \frac{\big[\frac{1}{2}\langle\rho\rangle_L u^{2}_{L}-\frac{3}{5}G\langle\rho\rangle_L\frac{M_L}{L}\big]}{L/u_L}={\rm inv}(L)~.
\end{equation}

In particular, it follows that the flux of kinetic energy per unit volume is also invariant:
\begin{equation}
 \label{eq_comp_turb_invariant}
 \langle \rho\rangle_L\,\frac{u_L^3}{L}={\rm inv}(L)~,
\end{equation}
in agreement with simulations of interstellar isothermal turbulence without gravity and with mixed forcing by \citet{Kritsuk_ea_07}.

\subsubsection{Diffuse zones outside the Perseus region}
\label{Diffuse_zones}
Mean density in the outer diffuse zones is found to scale with an index about the Larson's value (Fig. \ref{fig_n_scaling_cloud}, bottom). The consideration of their union ${\rm d}1\,\cup\,{\rm d}2\,\cup\,{\rm d}3$ yields the same result with a larger dispersion. We attribute the different scaling law in comparison with the PR rings to the actual variety of distances. All spatial scales and, respectively, mean densities in the diffuse zones outside the PR were calculated adopting $D=260$~pc (to the Perseus MC) which could be far from the real values for a number of lognormal components, in view of the large sizes of the zones and the remoteness of some parts of them from the cloud. 

\subsection{Comparison with numerical simulations}
\label{Comparison with simulations}
For a numerical test of the presented results, we ran a magnetohydrodynamical simulation of MC evolution including gravity performed with a {\sc FLASH} code \citep{Fryxell_ea_00}. The simulation properties are summarised in Table \ref{table_sim_summary}. The cloud formation was modelled through convergence of two large-scale cylindrical streams, each 112~pc long, with a radius of 64~pc. They are given an initial supersonic inflow velocity (isothermal Mach number of 2) in a warm neutral, initially homogeneous medium and collide at the centre of the numerical box. Turbulence is not driven continuously and, at late evolutionary stages, is due to fluid motions, driven by gravity. Feedback by sink particles is not included in the simulation.

\begin{table}
\caption{Summary of the used simulation.}
\label{table_sim_summary} 
\begin{center}
\begin{tabular}{lc}
\hline 
\hline 
% Zone & $N_{\rm max}$ & $\sigma_N$ & $N_{\rm low}^{\rm PL}$ & $n$ & $f_{\rm PL}(S)$ & $f_{\rm PL}(M)$\\ 
% \hline 
Box size & 256 pc \\	
Simulation code	& {\sc FLASH} \\	
Boundary conditions & periodic  \\	
Initial temperature & $T=5\,000$~K\\	
Initial density	& 1~\cc\\	
Turbulence & decaying \\	  
Mach number & 0.4\\	
Magnetic field (aligned along the X axis) & 3 $\mu$G \\
Refinement levels & 11 \\
Maximum resolution & $0.03$~pc\\ 
Jeans length & 8 grid cells \\
\hline 
\hline 
\end{tabular} 
\end{center}
\smallskip 
\end{table}

\begin{figure*} 
\begin{center}
\includegraphics[width=0.9\textwidth]{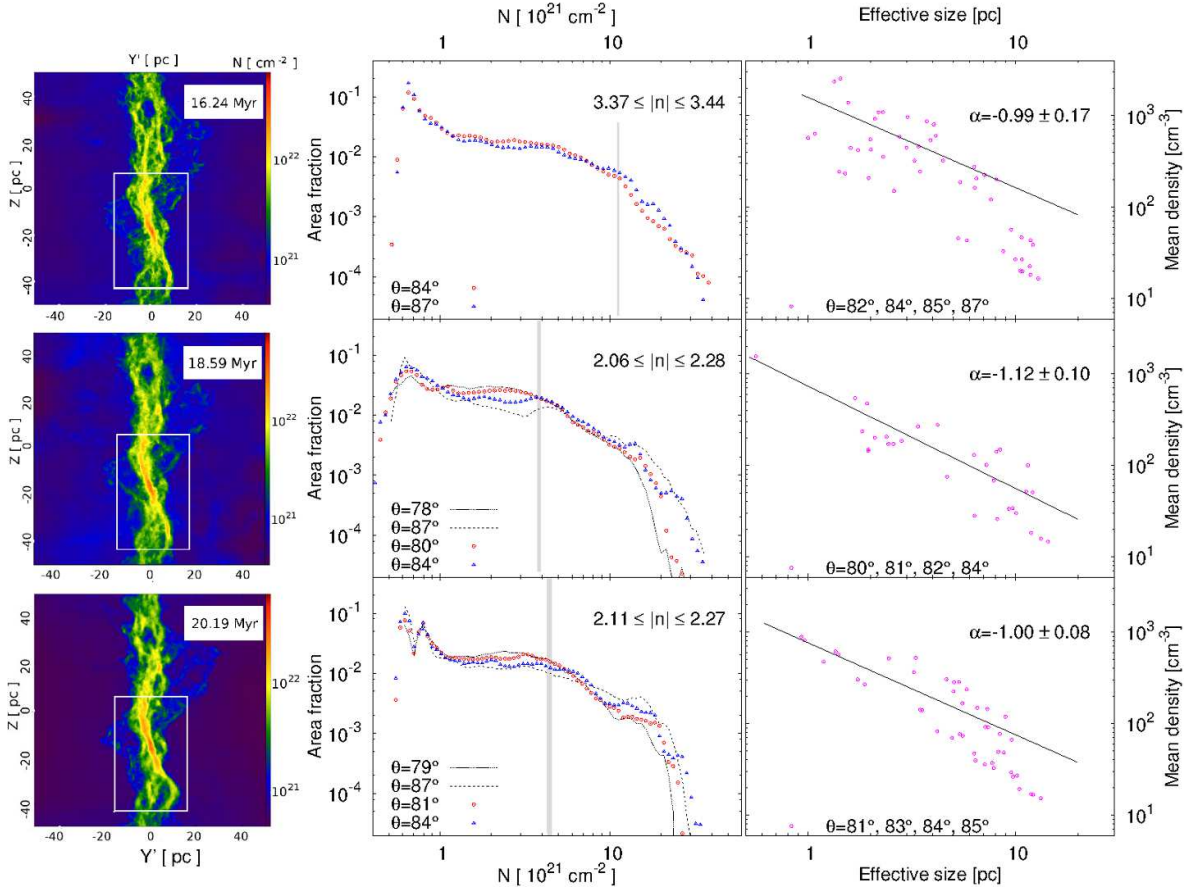}
\vspace{0.2cm}  
\caption{Column-density pdfs (middle) and scaling laws of mean density (right) at three evolutionary stages of MC evolution (left). All column-density maps are produced for a fixed line-of-sight angle $\theta=84^\circ$ and the considered rectangular region is indicated with solid white line. Shaded areas in the middle panel denote the location of the lower boundary $N_{\rm low}^{\rm PL}$ of the PL-tail. The pdfs for critical values of $\theta$ are plotted with dashed and dotted lines (see text).}
\label{fig_maps_PDF_scaling}
\end{center}
\end{figure*}

Similar numerical set-up was chosen in the works of \citet{VS_ea_07} and \citet{Banerjee_ea_09}. The former study showed the establishment of equipartition between gravitational and kinetic energy (cf. our physical assumption (ii) in Sect. \ref{Assumptions}) in dense gas domains ($>50$~\cc) and at evolutionary stages $t>10$~Myr. The latter work revealed that such condensations form mainly in knots of intersecting filamentary structures and those located in the central region of the flow collision plane host sink particles. Now, if one rotates the flow collision plane from a line-of-sight angle $\theta=0^\circ$ (face-on view) to $\theta\gtrsim80^\circ$, it should be expected that a cloud similar to Perseus MC will be ``observed'': an elongated, quasi-axisymmetric dense region with some filamentary `trunks' in its near vicinity (cf. our Fig. \ref{fig_probe_regions} with Fig. 1, right in \citealt{Banerjee_ea_09}). Indeed, this is the picture on the column-density maps obtained from our simulation at 
evolutionary stages $t>15$~Myr, plotted in Fig. \ref{fig_maps_PDF_scaling}, left.  

\subsubsection{Column-density pdfs}
We delineated a rectangular region of the same effective size like the PR (Zone \#6 in Table \ref{table_ezones_param}) and derived the $N$-pdfs in it, varying the line-of-sight angle in the range $70\le \theta \le88^\circ$. Some of the obtained distributions are shown in Fig. \ref{fig_maps_PDF_scaling}, middle. The PL-tail range as detected by {\sc Plfit} increases to about one order of magnitude at $t>18$~Myr. The slope $|n|$ gets shallower with evolution but stabilizes gradually at values $2.1-2.3$ for $80\lesssim\theta\lesssim84^\circ$, in excellent agreement with our result for the PR (Fig. \ref{fig_N-pdf_cloud}). Below some critical value $\theta\simeq79^\circ$ the PL tail becomes very steep and too short, i.e. $N_{\rm low}^{\rm PL}$ increases discontinuously by a factor 2-3. On the other hand, nearly side-on views ($\theta\gtrsim85^\circ$) are characterised by much longer and shallower PL tails: $|n|$ is close to unity. 

Considering only the non-PL part of the pdf ($N<N_{\rm low}^{\rm PL}$), we decomposed it to lognormal components through the procedure described in Sect. \ref{Fitting_procedure}. All obtained fits in the selected region, at three different evolutionary stages, are plotted in the Appendix \ref{Appendix_decomposition}, Fig. \ref{fig_Panel_Npdf_sim}, while the achieved fits' goodness is given in Table \ref{table_fit_goodness_sim}. We derived the mean-density scaling law, adopting the definition of $\langle \rho\rangle_L$ from Sect. \ref{Assumptions}. The results for the corresponding late evolutionary stages are illustrated in Fig. \ref{fig_maps_PDF_scaling}, right. 

Evidently, the mean density scaling index $\alpha\simeq-1$ does not evolve within the considered time-frame as the uncertainty decreases. Following the approach from Sect. \ref{Rings_in_Perseus_region} and \ref{Diffuse_zones}, one obtains for the mass and velocity scaling indices $\gamma\simeq2.0$ and $\beta\simeq0.5$, respectively. Hence, the total energy per unit volume turns out to be scale-invariant:
\begin{equation}
 \label{eq_invariant_simulation}
 \frac{1}{2}\langle\rho\rangle_L u^{2}_{L}-\frac{3}{5}G\langle\rho\rangle_L\frac{M_L}{L}={\rm inv}(L)~,
\end{equation}
while the kinetic energy flux per unit volume scales weekly $\langle \rho\rangle_L u_L^3/L\propto L^{-1/2}$. This behaviour is closer to the expected one for compressible turbulent medium with purely solenoidal forcing \citep{GB_11} but demonstrates that gravity nevertheless affects gas dynamics. The result might be affected as well by the presence of magnetic fields in our simulation. 

\subsubsection{Volume density pdfs}
\begin{figure} 
\begin{center}
\includegraphics[width=83mm]{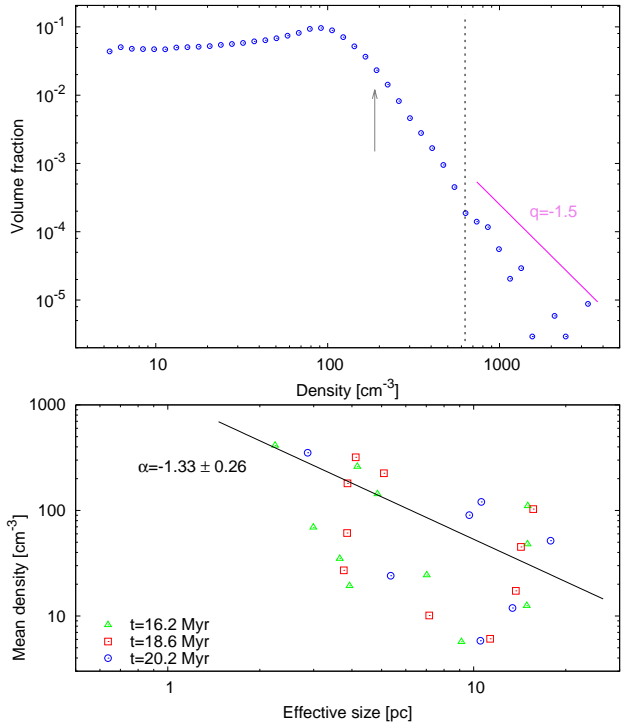}
\vspace{0.2cm}  
\caption{Pdf and scaling of volume density in the rectangular region delineated in Fig. \ref{fig_maps_PDF_scaling}, left. {\it Top:} Density distribution at $t=20.2$~Myr ($\theta=84^\circ$) with the lower limit of the PL tail from the {\sc Plfit} procedure (arrow) and the presumed actual one (dashed). The presumed slope of the PL tail is shown for reference. {\it Bottom:} Scaling law of mean density, averaged over three evolutionary stages.}
\label{fig_rho-PDF_scaling}
\end{center}
\end{figure}

The use of a numerical simulation allows for a direct study of the volume-density pdfs which correspond to those of column-density. An example of such $\rho$-pdf of the gas phase with densities $>5$~\cc~is shown in Fig. \ref{fig_rho-PDF_scaling}, top. The {\sc Plfit} procedure derives a very steep PL tail ($q\sim-3.7$) spanning about an order of magnitude and starting close to the distribution maximum. However, the statistics at high densities is very poor: a few dozens out of 175000 pixels have densities over 800~\cc,  which are typical for dense regions in MCs. High-resolution simulations of self-gravitating turbulent media \citep[e.g.][]{Collins_ea_11, KNW_11} produce $\rho$-pdfs whose PL tail starts at least $1.5$ orders of magnitude from the peak. Obviously, the PL tail of the $\rho$-pdf has not been resolved in our simulation. One is not able to derive the actual slope but the data hint at a value not far from $q=-1.5$ as shown in Fig. \ref{fig_rho-PDF_scaling}, top. Referring the reader again to the 
relations \ref{eq_Sigma-N_PDF} and \ref{eq_rho-rho_PDF}, note that $q\simeq-1.5$ yields a PL-tail slope of the $N$-pdf $n\simeq-2$ and $\alpha=-2$ which is consistent with the results for the PR (Figs. \ref{fig_N-pdf_cloud} and \ref{fig_n_scaling_cloud}, top) as well with the simulational $N$-pdfs (Fig. \ref{fig_maps_PDF_scaling}, middle).
 
Ignoring the presumed PL-tail (at densities $\gtrsim600$~\cc), we applied the procedure of lognormal decomposition (Sect. \ref{Fitting_procedure}) to the rest of the $\rho$-pdf. (The results are given in Appendix \ref{Appendix_decomposition}.) Thus the notion of spatial scales, introduced in Sect. \ref{Components_and_scales}, was retained as well in the 3D case; with the necessary correction of the coefficient in equation \ref{eq_eff_size_scale} to $(a_i/\sum_i a_i)^{1/3}$. The peak value $\rho_i$ of each lognormal component was adopted as mean density of the corresponding spatial scale. The number of lognormal components at each of the considered three evolutionary stages $16.2\le t \le 20.2$~Myr is high enough to allow for search of a correlation between density and size. The derived mean density scaling law is shown in Fig. \ref{fig_rho-PDF_scaling}, bottom. The slope varies between $\sim-1.0$ and $\sim-1.4$ but with a large scatter. Nevertheless, the correlation is seemingly independent on the 
evolutionary time although the spatial scales change their effective sizes and mean densities. Therefore it is safe to assume and to derive a single, time-averaged density scaling law, taking into account all data. The obtained slope $-1.33\pm0.26$ is steeper but -- in view of the large data scatter -- in general agreement with the result from the simulational $N$-pdf analysis.

\section{Two domains in the Perseus region: a hypothesis}
Our analysis hints at the existence of different physical regimes in two spatial domains associated with the PR: gravoturbulent in the Perseus MC and its vicinity, described by equation \ref{eq_invariant_MC}, and predominantly turbulent in the diffuse neighbourhood, described by equations \ref{eq_invariant_rings} and \ref{eq_comp_turb_invariant}. Below we propose an idea to estimate the effective size $L_{\rm gt}$ of the gravoturbulent domain and of the transition zone between the domains. 

Let $L$ and ${\cal L}$ are the scaling variables in the gravoturbulent and in the predominantly turbulent domain, respectively. For simplicity, the transition zone is considered as an extended shell encompassing the gravoturbulent domain, with effective size ${\cal L}_{\rm tr}$, i.e. with thickness $d_{\rm tr}=\sqrt{L_{\rm gt}^2+{\cal L}_{\rm tr}^2}-L_{\rm gt}$. The velocity dispersion $v_{\cal L}$ in this diffuse shell obeys the invariant flux of kinetic energy (equation \ref{eq_comp_turb_invariant}) and the mean density scales with index $\alpha\simeq-0.80$ (see Sect. \ref{Rings_in_Perseus_region}). On the other hand, adopting $\gamma\simeq1$ and $\beta\simeq0$ in the gravoturbulent domain (see Sect. \ref{PL_tail_zones}), one gets a linear scaling of mass $M_L=M_0 (L/1~{\rm pc})$ and a lack of scaling of velocity $u_L={\rm const}(L)$ there.

It is safe to assume that the time of consideration $dt$ is very small in comparison to the turbulent crossing time $\tau_{\cal L}$ in the predominantly turbulent domain and, accordingly, the mass increase $dM_L$ of the gravoturbulent domain is negligible \citep[][Sect. 2.4.1]{HF_12}. A further reasonable approximation is that the mass inflow is stationary: $dM_L/dt = {\rm const}(t) > 0$ (Klessen \& Hennebelle 2011). Then the derivative of the total energy $E_{\rm tot} = M_{L}u_{L}^{2}/2 - 3GM_{L}^{2}/5L$ is:
\begin{equation}
 \label{eq_mec_energy_derivative} 
  \frac{dE_{\rm tot}}{dt}=\frac{1}{2}u_L^2\frac{dM_L}{dt} - \frac{6}{5}G\frac{M_L}{L}\frac{dM_L}{dt} + \frac{3}{5}G\frac{M_L^2}{L^2}\frac{dL}{dt}~.
\end{equation}
where the first two terms in the right-hand side account straightforwardly for mass accretion and the last one -- for the size change associated with the mass increase. 

In view of the negligible $dM_L$, the main contributor to $dE_{\rm tot}/dt$ is the derivative of the kinetic energy of the predominantly turbulent domain: $|dE_{\rm tot}/dt|\approx \frac{1}{2}v_{\cal L}^2 |dM_{\cal L}/dt|$. Now, replacing $dL/dt=(1/M_0)dM_L/dt$ in equation \ref{eq_mec_energy_derivative} and using $|dM_L/dt|\approx|dM_{\cal L}/dt|$, one obtains:
\[  \frac{1}{2}v_{\cal L}^2=\big|\frac{1}{2}u_L^2 - \frac{3}{5}G\frac{M_L}{L}\big|=\big|(f-1)\frac{3}{5}G\frac{M_0}{1~{\rm pc}}\big|~, \]

\noindent recalling our basic assumption (ii) for equipartition $E_{\rm kin}=f|W|$ at each scale. From the derived mean density scaling law in the PR (Fig. \ref{fig_n_scaling_cloud}, blue symbols) one calculates directly $M_0=M_{L=1~{\rm pc}}=\langle\rho\rangle_{L=1~{\rm pc}}\,.{\rm pc}^3\sim10^3~M_\odot$. Thus we arrive at an expression to estimate ${\cal L}_{\rm tr}$, assessing the coefficient $f$ in the gravoturbulent domain and the coefficient in the velocity scaling law $v_{\cal L}=v_0 ({\cal L}/1~{\rm pc})^\beta$ in the predominantly turbulent domain:
\begin{equation}
 \label{eq_boundary_zone_estimate}
 \Big(\frac{{\cal L}_{\rm tr}}{1~{\rm pc}}\Big)^{2\beta}=\frac{6}{5}\frac{|f-1|}{v_0^2}G\frac{M_0}{1~{\rm pc}}~,~~~\beta\simeq0.60
\end{equation}

\begin{figure} 
\begin{center}
\includegraphics[width=83mm]{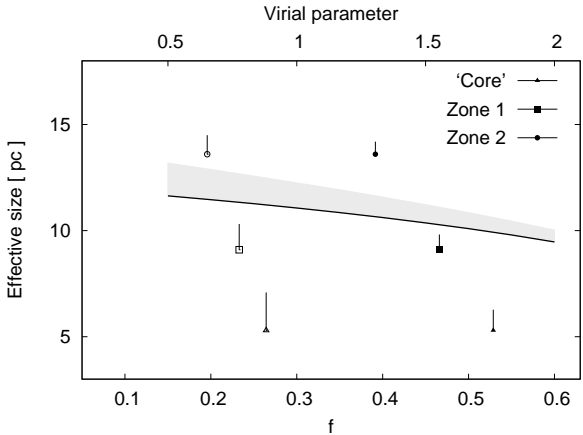}
\vspace{0.2cm}  
\caption{Modelled sizes of the gravoturbulent domain $L_{\rm gt}$ (solid line) and of the transition zone $d_{\rm tr}$ (grey area) as functions of $f$, fixing $v_0=0.8$~km/s and $\beta=0.60$ in the turbulent domain. The locations of the `Core', Zone 1 and Zone 2 in the limiting cases of $f$ (minimal value - open symbols; maximal value - filled symbols) are calculated according to Fig. 1 in \citet{KPG_13}; the bars show the extension of the transition zones. See text.}
\label{fig_trans_zone}
\end{center}
\end{figure}

Typical values of $f$ and $v_0$ in star-forming regions are provided from the extensive recent study of the virial parameter $\alpha_{\rm vir}$ in MCs and embedded MC fragments by \citet{KPG_13}. Those authors define $\alpha_{\rm vir}=g(p)\,2E_{\rm kin}/|W|$, where $g(p)$ is a geometrical factor depending on the density gradient with index $p$. They found $0.2\lesssim\alpha_{\rm vir}\lesssim2$ and $0.7\lesssim v_0\lesssim1.1$~km/s for the majority of sampled objects with masses $10^3$ to few times $10^4$\Msol~which correspond (through $M_0$) to spatial scales $1-30$~pc considered in this Paper. For a variety of possible density profiles an appropriate mean value of the geometrical factor is $g(p)\simeq1.4$ \citep{BM_92} and hence $0.07\lesssim f \lesssim 0.7$, depending on the mass of the MC fragment. 

Assuming that mass in the transition zone still scales linearly, like in the gravoturbulent domain, one gets an equation for $L_{\rm gt}$: 
\begin{equation}
 \label{eq_L_gravoturbulent}
 M(d_{\rm tr})=M_0 d_{\rm tr} \propto {\cal L}_{\rm tr}^\alpha \big[ (L_{\rm gt}+d_{\rm tr})^3 - L_{\rm gt}^3 \big]\,,~\alpha\simeq-0.80~~
\end{equation}

The obtained modelled values of $L_{\rm gt}$ and $d_{\rm tr}$ as functions of $f$ (or of the virial parameter), for a fixed $v_0=0.8$~km/s, are shown in Fig. \ref{fig_trans_zone}. Making use of the observational correlation between virial parameter and mass \citep[][Fig. 1]{KPG_13} and of linear scaling of mass $M_L=M_0 (L/1~{\rm pc})$, one can compare the prediction of the proposed model with the locations of the internal elliptical zones of the PR. Evidently, the gravoturbulent domain is somewhat larger than Zone \#1 which encompasses Perseus MC as a compact and apparently axisymmetric region. This domain determines the $N$-pdf characteristics (Fig.~\ref{fig_N-pdf_cloud}, right) and the density scaling law (Fig.~\ref{fig_n_scaling_cloud}) in the PR considered as a whole, while these quantities in the set of rings are typical for diffuse, predominantly turbulent regions. It is interesting also to note that the column-density range $N\gtrsim10^{21}$~\sqc~in the gravoturbulent domain (see the pdf of Zone \#1 in Fig. \ref{fig_N-pdf_cloud}, left) corresponds entirely to the molecular gas phase in Perseus according to the study of \citet{Lee_ea_12}.

\section{Discussion}
\subsection{Possible uncertainties of the density scaling law}

The derivation of density scaling laws could be sensitive to various factors. Here we examine briefly three of them: map resolution, uncertainties due to the distance gradient to the PR and uniqueness of $N$-pdf decomposition. Our results were obtained by use of the original map resolution which corresponds to $\sim 0.1$~pc, i.e. about the size of dense clumps. We mimicked the effect of lower resolution using moving median smoothing of the column-density map, as described in Appendix \ref{Appendix_decomposition} wherein the corresponding change in the pdf shape is illustrated (Fig. \ref{fig_N-pdf_rings_smoothed}). Although one or two additional components would be required in such a case to improve fit's goodness, the derived slopes of the density scaling law (Fig. \ref{fig_n_scaling_cloud}) practically do not change -- both in the embedded zones and in the diffuse rings. The distance uncertainties in the whole PR can be assessed roughly since the estimates of \citet{Schlafly_ea_14} cover only the cloud itself and its close vicinity: Zones \#1 and \#2. Linear fitting of those data yields an approximately constant distance gradient whose vector is nearly collinear to the large axis of the elliptical regions. Therefore one may consider the whole region as being rotated at an approximately constant angle $\psi$ around the small axis. Such geometry would affect the effective radii $R$ and sizes $L_i$ (see eq. \ref{eq_eff_size_scale}) by factor $\cos\psi$ and, respectively, would shift the zero point of the derived density scaling laws (Fig. \ref{fig_n_scaling_cloud}) but would {\it not} affect the scaling indices.

Next, one might question the uniqueness of $N$-pdf decomposition to a small number of lognormals. Is it possible to fit the observational column-density distribution by sum of another set of lognormals which shows up different physical features? Such arbitrariness of the suggested approach is severely limited through its starting steps: 1) setting the number of components equal to the number of prominent local peaks of the $N$-pdf and, 2) choosing such guess values of $(N_i,\,\sigma_i,\,a_i)$ which nearly reproduce the location of each peak and the $N$-pdf shape in its vicinity. Finally a simultaneous Levenberg-Marquardt optimisation of all $3{\cal N}$ parameters is performed to obtain the fitting curve. On the other hand, the termination of the fitting procedure depends on estimation of the fit goodness -- in principle, the latter can be improved by adding one or two further components to account for single tiny fluctuations of the 
distribution. However, such additional component(s) would be of small effective size (small $a_i/\Sigma_i a_i$) in comparison with the bulk of components and could slightly affect the scaling of mean density (cf. Fig. \ref{fig_n_scaling_cloud}). The defined statistical weights (equation \ref{eq_components_weight}) also exclude essential changes of the mean density scaling index with adding up small components.

\subsection{Density scaling laws and physical regimes}
It may seem surprising that the derived mean-density scaling law in the diffuse rings $\alpha\simeq-0.80$ differs substantially from the one in the PR whereas -- expressed in terms of mass-size relationship ($\gamma\simeq2.2$, equation \ref{eq_gamma_alpha}), -- it resembles that of a large sample of Galactic MCs with sizes from a few to several tens pc studied by \citet{Roman-Duval_ea_10}. In fact, such behaviour is to be expected if one differentiates internal general structure of {\it individual} clouds from consideration of a sample of clouds. Total masses of MCs correlate well with their sizes following a single power law with $2\sim \gamma \lesssim 2.4$ \citep{LAL_10, Roman-Duval_ea_10} while the index $\gamma$ {\it within} a cloud increases reversely to the scale $L$. The latter phenomenon was discovered by \citet[][see Fig. 2 there]{LAL_10} from a dust extinction study of 11 Galactic MCs and explained within a statistical model of general MC structure \citep{DVK_11}, assuming equipartition of energies at each scale, characterized by its own pdf. The mean-density scaling law in the PR derived in this Paper confirms those results extending their applicability to the vicinity of the cloud itself (Fig. \ref{fig_n_scaling_cloud}; $L>10$~pc) while the value of $\gamma$ in the dense cloud regions (within the `Core') might increase. A further investigation by use of {\sc Planck} data, including other MCs with different general structure and various physical conditions (e.g. star-forming activity), would shed light on this important issue.

On the other hand, the diffuse rings evidently obey a mass scaling law like that for single MCs. The power-law index $\gamma \simeq 2.2$ is also in agreement, within the uncertainties, with the measurements from simulations \citep{Kritsuk_ea_07, FKS_09, Kritsuk_ea_09}. This result, combined with our assumption (ii) for equipartition between gravitational and kinetic energies, lends support to the hypothesis that supersonic compressible turbulence dominates the diffuse gaseous structures in cold interstellar medium as described by equations \ref{eq_invariant_rings} and \ref{eq_comp_turb_invariant}. It should be pointed out that assumption (ii) refers to abstract scales as defined in Sect. \ref{PDF_analysis_diffuse_zones} and {\it not} to connected regions with well-defined boundaries for which surface energy terms are comparable to the volume ones. Nevertheless, it is noteworthy that the mean-density scaling law of abstractly defined scales -- both from observational data on the PR and from the numerical simulations (Sect. \ref{Comparison with simulations}) -- are close, within uncertainties, to $\alpha\simeq-1.1$ derived by \citet{Larson_81} for MCs and cloud fragments delineated as connected regions. Moreover, the striking coincidence between the derived scaling indices from column- and volume-density pdfs from our simulation confirms this result and supports the assumption of isotropy in turbulent gas medium. Thus the so defined spatial scales reproduce the fractal MC structure as derived by use of classical clump-finding techniques.

The mean-density scaling law $\alpha\simeq-1$ obtained from the performed simulations is slightly steeper than the one in the diffuse rings in the PR and seems to indicate slightly different physical conditions. One may attempt to integrate the obtained results for the PR and those from the simulations in a single scenario, interpreting the variety of physical invariants and scaling laws as referring to different evolutionary stages. A clear indication for this is the behaviour of the flux of kinetic energy per unit volume $\langle\rho\rangle_L u_L^3/L$ (equation \ref{eq_kinetic_energy_flux}). The latter quantity is scale-independent in the diffuse rings of the PR, scales as $L^{-1/2}$ in the diffuse components of the simulational pdfs and scales as strongly as $L^{-3}$ in the elliptical zones of the PR. That points to the increasing role of gravity in the gas dynamics. Possibly, the assumed three physical regimes hint at different self-similar gas structure.

\subsection{Relation to the velocity scaling law}
The density scaling laws in the PR are derived only by use of the procedure for $N$-pdf decomposition, without any reference to the velocity scaling law. The relation between the two scaling laws relies essentially on our second physical assumptions for equipartition between gravitational and kinetic energy (Sect. \ref{Assumptions}). Introducing the latter, we aimed to explore the physics which governs the column-density statistics in star-forming regions and {\it not} to derive velocity-size relationships out of column density data. Nevertheless, a comparison of the (indirectly) estimated scaling index $\beta$ with observational and numerical studies might be instructive. 

\begin{figure} 
\begin{center}
\includegraphics[width=83mm]{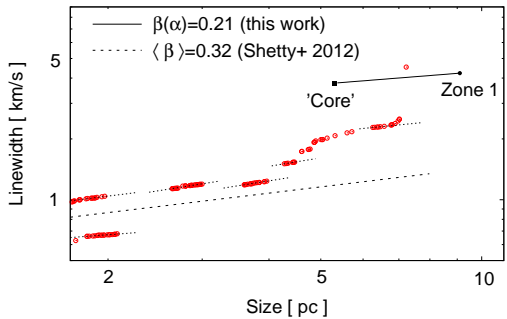}
\vspace{0.2cm}  
\caption{Velocity-size diagram of large embedded (dendrogram) structures in the Perseus region studied by \citet{Shetty_ea_12}, with their unique scaling laws (dotted). The index $\beta$ in the range of overlapping with the present Paper is estimated through the density scaling index $\alpha$ and shown for comparison.}
\label{fig_velocity-size}
\end{center}
\end{figure}

The linewidth-size relations in the Perseus MC, obtained by \citet{Shetty_ea_12} from \thirthCO COMPLETE survey data, are plotted in Fig. \ref{fig_velocity-size}. The considered objects are hierarchical clumps, extracted by use of the {\sc Dendrogram} algorithm \citep{Rosolowsky_ea_08}. Therefore they are very appropriate for comparison to our embedded elliptical zones. Unfortunately, their size range overlaps with our study only within Zone \#1 and its `Core' (cf. Fig. \ref{fig_n_scaling_cloud}, up) so that we are able to present only a qualitative analysis here. A dendrogram of clumps is a multilevel hierarchy in which large embedded structures as well single condensations are distinguished (see Fig. 1 in \citealt{Shetty_ea_12}). It has been discovered from analysis of simulations of self-gravitating turbulent medium that series of massive dendrogram structures are characterized by unique mass-density and velocity-size relationships \citep{Stanchev_ea_13}. Such structures and their unique velocity-size relationships are shown in Fig. \ref{fig_velocity-size} with short dotted lines. They are representative for scales $L\gtrsim2$~pc within the Perseus MC. The averaged scaling index $\langle \beta \rangle$ is in a rough agreement with the estimation from results in this Paper for the `Core' and Zone \#1. A tendency of shallowing in regard to the embedded zones outside Perseus MC is sensible, although the lack of molecular-line data for larger scales leaves the issue open.  

The simulation used in this Paper to compare with the derived density-scaling laws is not relevant to test the velocity-size relationship. It implements decaying initial turbulence which cannot provide a large inertial range to derive the velocity scaling law, although accretion on protostellar objects might be an effective driving mechanism at small scales. Nevertheless, the initial turbulence shapes significantly the density field in earlier evolutionary stages and hence is crucial for the cloud morphology.

The virtual lack of velocity scaling in the PR (equation \ref{eq_no_scaling_velocity}) indicates the essential role of gravity. It is noteworthy that the estimated effective size of the gravoturbulent domain is about the size of the Perseus MC (cf. Table \ref{table_ezones_param} and Fig. \ref{fig_trans_zone}) for a variety of conditions: from weak to strong gravitational boundedness ($0.25\lesssim f \lesssim0.5$). The size $d_{\rm tr}$ of the transition zone between gravoturbulent and predominantly turbulent regimes depends mainly on the balance between $E_{\rm kin}$ and $|W|$ in the former regime and of coefficient $v_0$ of the velocity scaling law in the latter one. The typical values of $d_{\rm tr}$ are about 10 \% of size of the gravoturbulent domain, irrespective of the degree of gravitational boundedness.

\subsection{The role of magnetic fields}
Finally, we briefly consider the possible effect of magnetic fields which were not taken into account in the adopted physical frame (Sect. \ref{Assumptions}). Detailed comparison between cloud evolution from simulations in non-magnetized and in moderately magnetized medium shows that the magnetic support delays the gravitational collapse whereas the corresponding stages of the $N$-pdf evolution are identical, differing only by a small time factor \citep{Henne_ea_08}. About the time when gravitational collapse of the whole cloud is to start (i.e. when equipartition between gravity and turbulence is achieved), the shape of the column-density pdf in both cases is a combination of lognormal function plus power-law tail at $N\gtrsim$ few times $10^{21}$~\sqc~(see Fig. 1 in \citealt{Henne_ea_08}). Therefore one could not expect that the presence of magnetic fields would affect generally the presented physical analysis, given than the exact evolutionary stage of the PR is not known.

\section{Conclusions}
In this Paper, we presented an analysis of the column-density pdfs in the star-forming region Perseus and its diffuse environment using the dust opacity map at 353 GHz available from the {\sc Planck} archive. Due to its high galactic latitude and approximately axisymmetric form, free of contaminating fore- or background cloudlets, the region is an appropriate test case for studying the general structure of molecular clouds and the physical processes that govern their evolution. On the other hand, the {\sc Planck} data on dust emission offer the opportunity to probe regions of very high extinction, not resolved on dust extinction maps. 

Our main conclusions are as follows:
\begin{enumerate}
 \item[1.] The pdf shape can be fitted: i) by a combination of a lognormal function and an extended power-law tail at high densities, in zones centred at the molecular cloud Perseus; and, ii) by a linear combination of several lognormals, in rings surrounding the cloud or in zones of its diffuse neighbourhood. In the first case, the power-law tail of the pdf in the largest zone includes the pdf density ranges of all others (embedded) zones whose sizes may be considered as subscales of this zone. In the second case, each lognormal component of a given pdf is interpreted as a contribution of a separate spatial scale in the turbulent cascade. In that way, the notion of spatial scales $L$ is introduced which allows for derivation of power-law scaling laws of mean density $\langle \rho\rangle_L \propto L^\alpha$ and mass $M_L \propto L^\gamma$ in the studied regions.  
 \item[2.] The derived scaling laws in zones centred on the Perseus MC and those in zones of its diffuse vicinity or neighbourhood differ substantially: $\alpha=-1.93$ ($\gamma=1.07$) in the former case and $\alpha=-0.77\pm0.11$ ($\gamma=2.23$) in the latter case. Assuming an equipartition relation between gravitational and kinetic (turbulent) energy at each spatial scale and a power-law scaling of velocity dispersion $u_L\propto L^\beta$, this result bears evidence of two distinct physical regimes in the Perseus region:
 \begin{itemize}
  \item[-] gravoturbulent, characterized by nearly linear scaling of mass and practical lack of velocity scaling.
  \item[-] predominantly turbulent, characterized by a steep velocity scaling ($\beta\simeq0.60$) and by invariant for compressible turbulence $\langle\rho\rangle_L u_L^3/L$, describing a scale-independent flux of the kinetic energy through turbulent cascade.
  \end{itemize}
  General identification of these regimes with spatial domains in the Perseus region is physically sustained in point 5 bellow. 
  \item[3.] The mean-density scaling law $\alpha\simeq-1$, derived both from column- and (within the a large data scatter) volume-density pdfs from the performed magnetohydrodynamical simulations with gravity, is in excellent agreement with the classical study of \citet{Larson_81}. This is striking since the result was obtained for abstractly defined spatial scales -- not for connected regions in MCs, delineated by use of some clump-finding technique.   
 \item[4.] The obtained three different physical invariants (resp., scalings of $\langle\rho\rangle_L u_L^3/L$) in the PR, its diffuse rings  and in the simulated cloud region seem to point to three different types of self-similar structure.
 \item[5.] Modelling the gravoturbulent spatial domain as centrally symmetric, with effective size $L_{\rm gt}$, and the predominantly turbulent domain as its extended shell, we estimate $L_{\rm gt}$ and the thickness of the transition zone $d_{\rm tr}$. For reasonable values of the virial parameter, one obtains $L_{\rm gt}\sim10-11$~pc and $d_{\rm tr}\sim0.6-1.5$~pc, i.e. the gravoturbulent domain can be generally identified with the Perseus MC in whose vicinity a relatively sharp transition occurs to predominantly turbulent physical regime. The latter roughly coincides with the transition from molecular to atomic gas phase \citep{Lee_ea_12}. 
\end{enumerate}
\vskip12pt
{\it Acknowledgement:} The data used in this work are from observations obtained with Planck (http://www.esa.int/Planck), an ESA science mission with instruments and contributions directly funded by ESA Member States, NASA, and Canada. The simulations used in this work were run on HLRN-III (Hannover, Germany) under project grant hhp000222.\\ T.V. acknowledges support by the {\em Deutsche Forschungsgemeinschaft} (DFG) under grant KL 1358/15-1. R. S. K. thanks for the subsidies from the DFG in subprojects B1, B2 and B5 of SFB 881 ``The Milky Way System'', as well as support via the priority program SPP 1753 ``Physics of the Insterstellar Medium''.

% \newpage

\label{lastpage}

\newpage
\appendix
\section{Decomposition of the pdf: lognormal components and fit goodness}
\label{Appendix_decomposition}
\begin{figure} 
\begin{center}
\includegraphics[width=84mm]{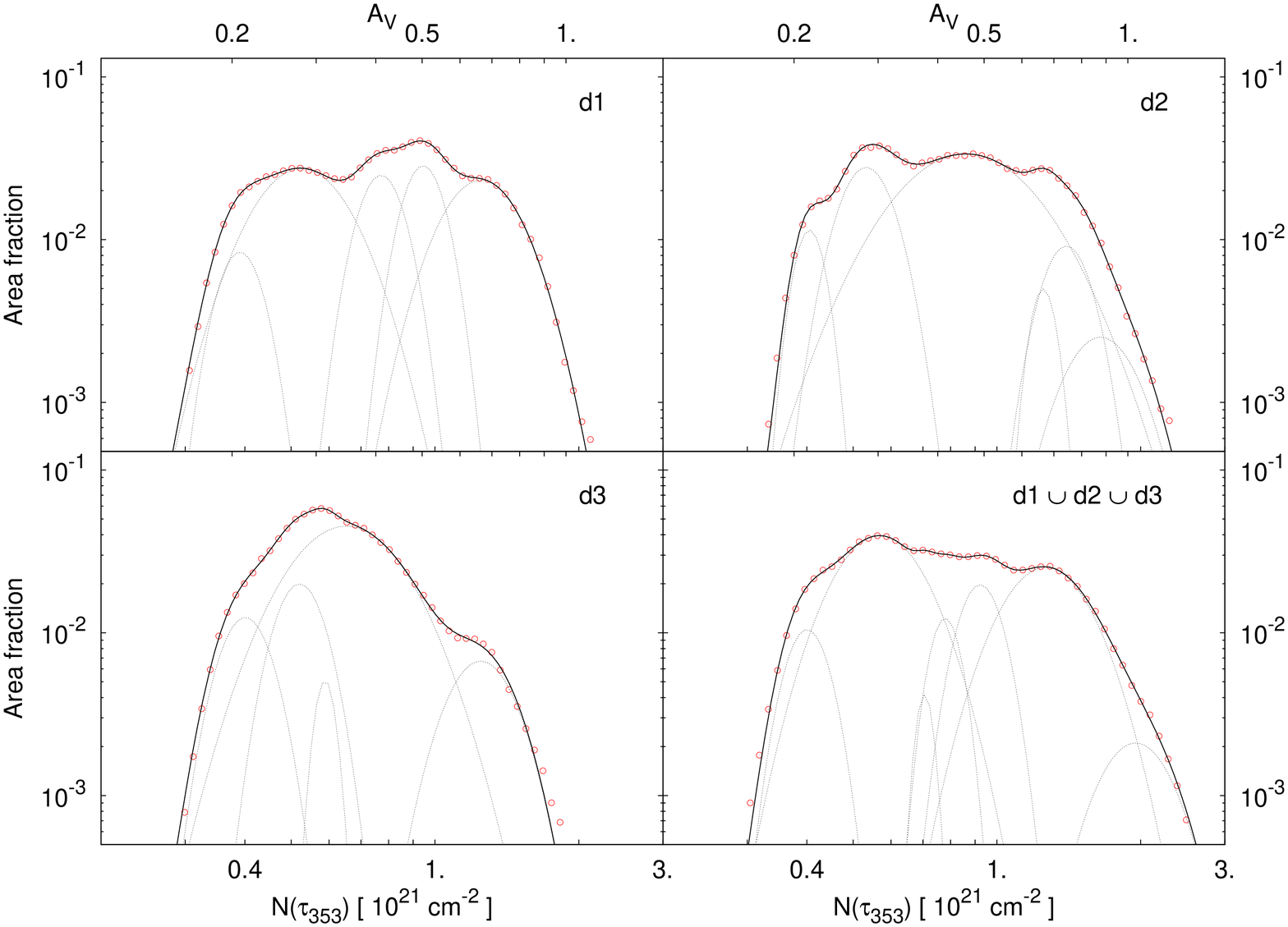}
\vspace{0.3cm}  
\caption{Decomposition of the $N$-pdf (open symbols) in the diffuse zones to lognormals (dotted). The fitting function is plotted with solid line.}
\label{fig_N-pdf_dif}
\end{center}
\end{figure}

\begin{figure} 
\begin{center}
\includegraphics[width=84mm]{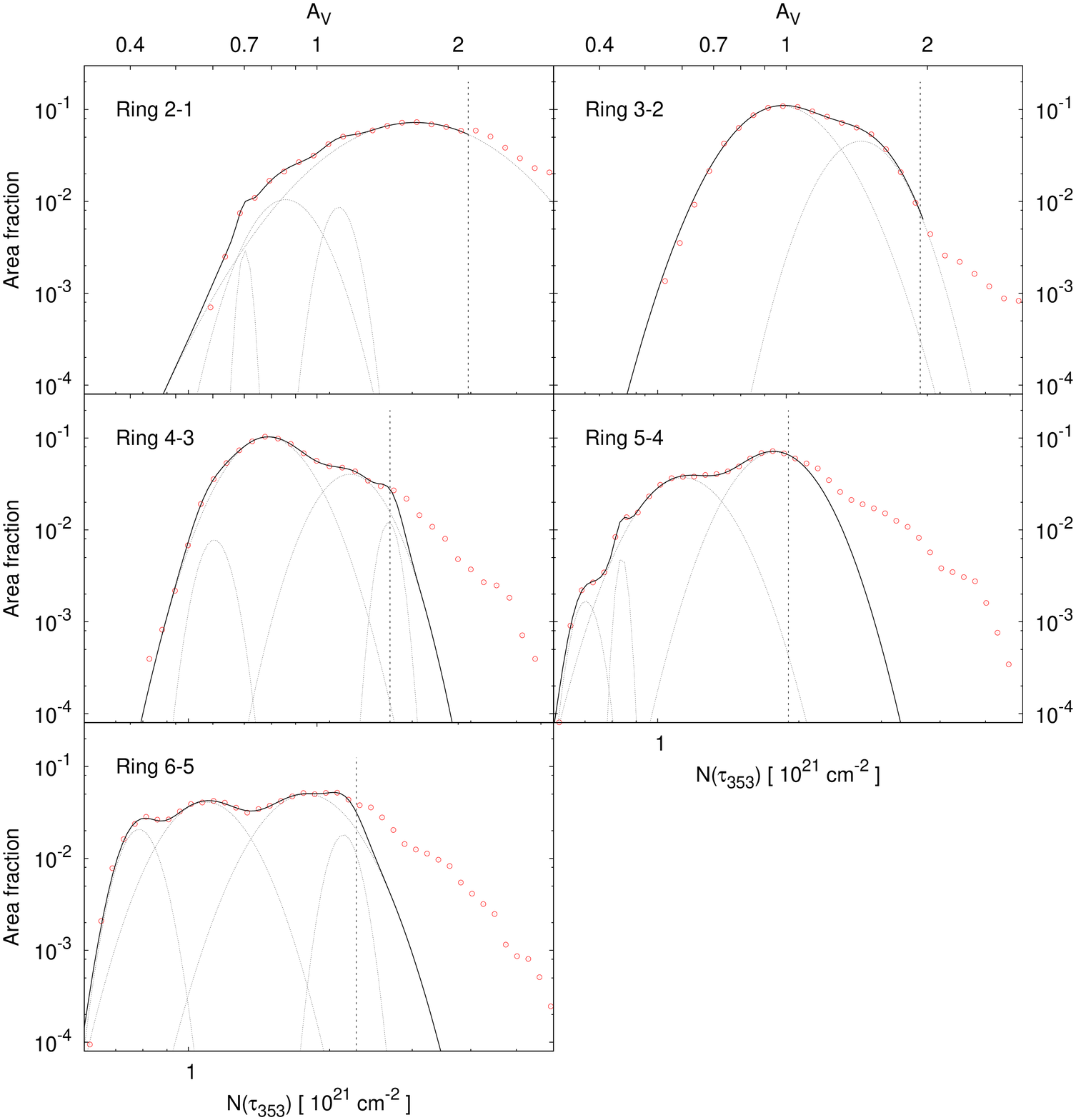}
\vspace{0.3cm}  
\caption{Decomposition of the $N$-pdf (open symbols) to lognormals in rings delineated by the boundaries of two successive elliptical zones in the PR (cf. Table \ref{table_ezones_param}). The locations of the lower boundaries $N_{\rm low}^{\rm PL}$ of the PL tail are indicated (dashed). Other notation is the same like in Fig. \ref{fig_N-pdf_dif}.}
\label{fig_N-pdf_rings}
\end{center}
\end{figure}

\begin{figure} 
\begin{center}
\includegraphics[width=84mm]{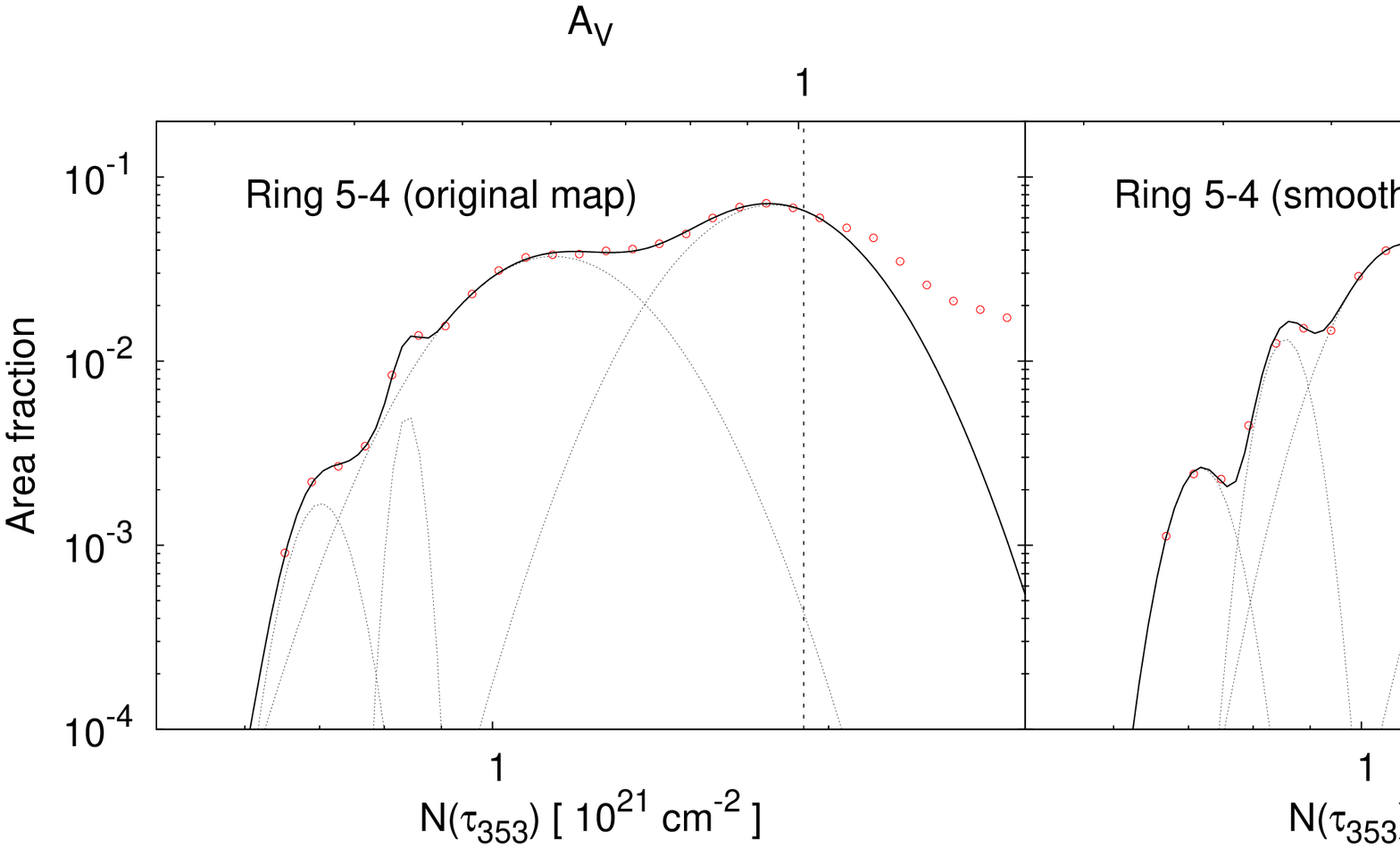}
\vspace{0.3cm}  
\caption{Illustration of the effect of map smoothing (mimicking lower resolution) on the pdf decomposition. See text.}
\label{fig_N-pdf_rings_smoothed}
\end{center}
\end{figure}

\begin{figure} 
\begin{center}
\includegraphics[width=.5\textwidth]{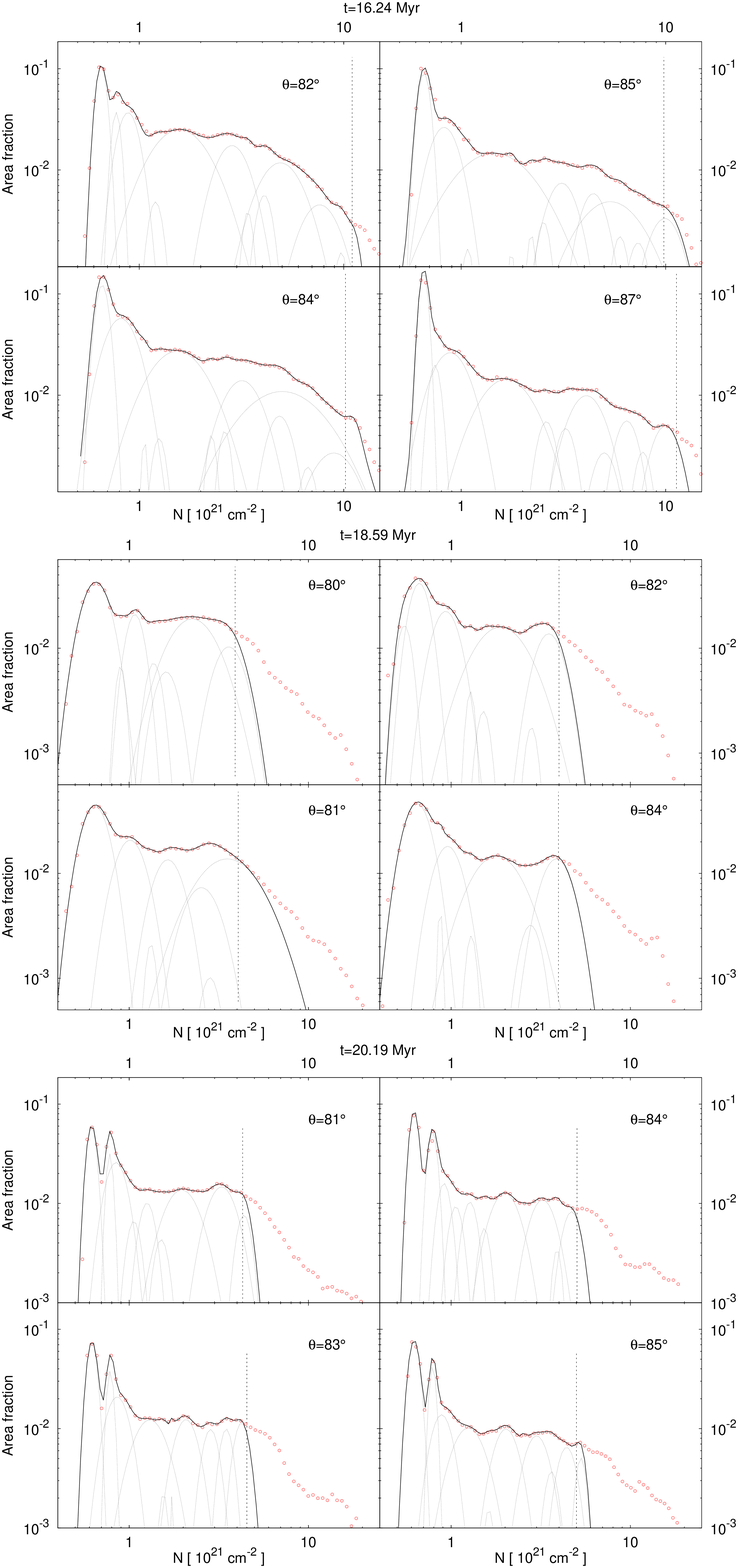}
\vspace{0.3cm}  
\caption{Decomposition of the $N$-pdf (open symbols) to lognormals in the simulated regions at three evolutionary stages of MC evolution (cf. Table \ref{table_fit_goodness_sim}).}
\label{fig_Panel_Npdf_sim}
\end{center}
\end{figure}

Below we show the pdf decomposition to lognormal components in the diffuse zones (Fig. \ref{fig_N-pdf_dif}) and in the rings outlined by the boundaries of embedded elliptical zones in the Perseus region (Fig. \ref{fig_N-pdf_rings}). For comparison, the pdf decompositions in the simulated region at three different evolutionary stages (cf. Fig. \ref{fig_maps_PDF_scaling}) are plotted in Fig. \ref{fig_Panel_Npdf_sim}. The goodness of the obtained fits of observational and numerical pdfs are given in Tables \ref{table_fit_goodness_obs} and \ref{table_fit_goodness_sim}, respectively. We studied the possible effect of insufficient resolution on the pdf, applying the technique of moving median smoothing on the original {\sc Planck} map: the value of $N$ in each pixel was replaced by the median value in a square containing the neighbor pixels. A typical result is shown in Fig. \ref{fig_N-pdf_rings_smoothed}. As expected, a lower map resolution would strengthen local peaks in $N$ (i.e. the densities with higher probability) -- the pdf becomes less smoothed and could require one or two additional components to achieve a satisfactory fit goodness. However, this additional components do not influence practically the derived density scaling law (slope changes of a few hundredths).

\begin{figure} 
\begin{center}
\includegraphics[width=53mm]{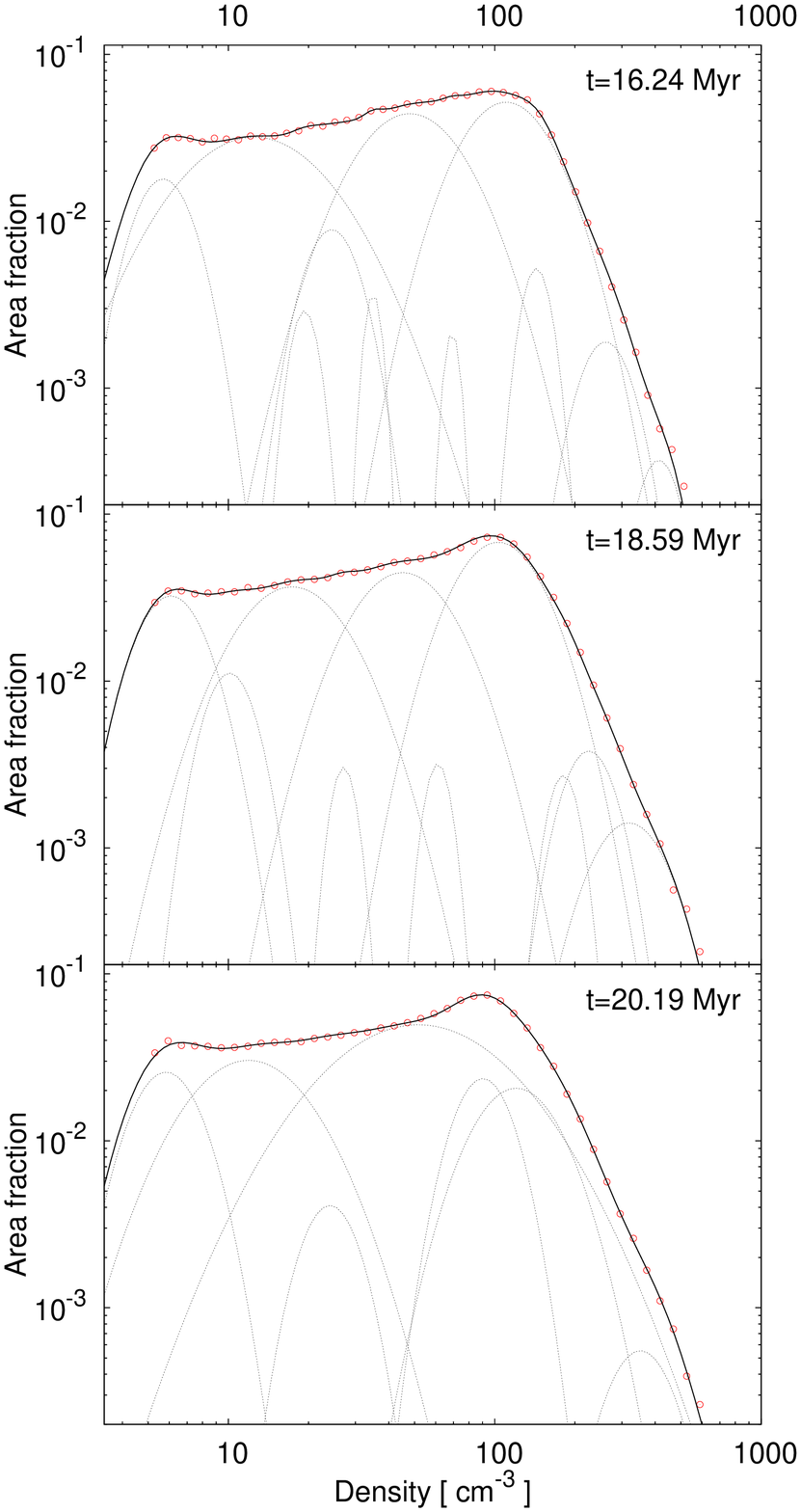}
\vspace{0.2cm}  
\caption{Decomposition of the volume-density pdf (open symbols) to lognormals in the simulated regions at three evolutionary stages of MC evolution (cf. Table \ref{table_fit_goodness_sim_rho}).}
\label{fig_pdf_panel}
\end{center}
\end{figure}

The decompositions of the volume-density pdf in the simulated region at three different evolutionary stages are plotted in Fig. \ref{fig_pdf_panel} and the goodness of the corresponding fits is given in Table \ref{table_fit_goodness_sim_rho}.

Evidently, the goodness of the obtained fits is satisfactory -- in most cases (except for Ring 3-2 in the PR and at $t=16.24$~Myr in the simulations), the $\chi^2$ value is consistent with the two-sided test at 95\% confidence level. As one can expect, the PL tail in simulational $N$-pdfs grows in terms of column-density range and statistical weight with the MC evolution (Fig. \ref{fig_Panel_Npdf_sim}). In contrast, the density-scaling law derived from the non-PL part does not change, with slope $\alpha \sim-1$ and uncertainty which decreases distinctly with evolution.

\newpage
\begin{table}
\caption{Goodness of the obtained fits to N-pdfs in the rings of the PR and in the diffuse zones. Notation: $N_{\rm bin}$ -- number of bins, $m$ -- total number of lognormal components, DF = degrees of freedom ($N_{\rm bin}-3m$), $\chi_{\rm crit,\,low\,(up)}^2$ - lower (upper) critical value of the $\chi^2$-distribution at 95\% confidence level. The $\chi^2$ values, which are consistent with the two-sided test $\chi_{\rm crit,\,low}^2\le\chi^2\le\chi_{\rm crit,\,up}^2$, are put in bold.} 
\label{table_fit_goodness_obs} 
\begin{center}
\begin{tabular}{cccccc@{~}c}
\hline 
\hline 
Region & $N_{\rm bin}$ & $m$ & DF & $\chi^2$ & $\chi_{\rm crit,\,low}^2$ &  $\chi_{\rm crit,\,up}^2$ \\ 
\hline 
\multicolumn{5}{l} {\it Rings in the PR} \vspace{4pt}\\
2-1 & 19 & 4	& ~7	& ~{\bf 2.99} & 2.17 & 14.07\\	
3-2 & 20 & 2	& 14	& ~1.58 & 6.57 & 23.69 \\	
4-3 & 22 & 4	& 10	& {\bf 16.91} & 3.94 & 18.31 \\	  
5-4 & 22 & 4	& 10	& ~{\bf 4.68} & 3.94 & 18.31 \\	
6-5 & 26 & 4	& 14	& ~{\bf 8.96} & 6.57 & 23.69\vspace{6pt}\\	
\multicolumn{5}{l} {\it Diffuse zones in the neighborhood} \vspace{4pt}\\
d1 & 55 & 5	& 40	& {\bf 42.53} & 26.51 & 55.76 \\	 
d2 & 51 & 6	& 33	& {\bf 34.96} & 20.87 & 47.40 \\	 	
d3 & 52 & 5	& 37	& {\bf 51.17} & 24.08 & 52.19 \\	
d1 $\cup$ d2 $\cup$ d3 & 60 & 7	& 39	& {\bf 51.80} & 25.70 & 54.57 \\	
\hline 
\hline 
\end{tabular} 
\end{center}
\smallskip 
\end{table}

\begin{table}
\caption{Goodness of the obtained fits to N-pdfs at three evolutionary stages of simulated MC evolution. The line-of-sight angle is denoted by $\theta$. Other notation is the same like in Table \ref{table_fit_goodness_obs}.} 
\label{table_fit_goodness_sim} 
\begin{center}
\begin{tabular}{cccccc@{~}c}
\hline 
\hline 
$\theta$ & $N_{\rm bin}$ & $m$ & DF & $\chi^2$ & $\chi_{\rm crit,\,low}^2$ &  $\chi_{\rm crit,\,up}^2$ \\ 
\hline 
\multicolumn{7}{c}{$t=16.24$~Myr} \vspace{2pt}\\
82 & 58 & 12	& 22	& {\bf 31.70} & 12.34 & 33.92\\	
84 & 54 & 12	& 18	& {\bf 22.98} & ~9.39 & 28.87\\	
85 & 54 & 13	& 15	& {\bf 21.47} & ~7.26 & 25.00 \\	  
87 & 55 & 12	& 19	& {\bf 26.06} & 10.12 & 30.14 \vspace{2pt}\\	
\multicolumn{7}{c}{$t=18.59$~Myr} \vspace{2pt}\\
80 & 31 & ~8	& ~7	& ~1.55 & 2.17 & 14.07\\	
81 & 32 & 11	& 11	& {\bf 12.53} & ~4.58 & 19.68 \\	
82 & 32 & ~8	& ~8	& {\bf ~4.42} & ~2.73 & 15.51 \\	  
84 & 33 & ~8	& ~9	& {\bf 10.71} & ~3.33 & 16.92 \vspace{2pt}\\
\multicolumn{7}{c}{$t=20.19$~Myr} \vspace{2pt}\\
81 & 33 & 10	& ~3	& ~{\bf ~1.32} & ~0.35 & ~7.82\\	
83 & 34 & 11	& ~1	& ~{\bf ~0.98} & 0.01 & ~3.84 \\	
84 & 34 & 11	& ~1	& {\bf ~1.57} & 0.01 & ~3.84 \\	  
85 & 35 & 11	& ~2	& ~{\bf ~5.93} & 0.10 & ~5.99 \\	
\hline 
\hline 
\end{tabular} 
\end{center}
\smallskip 
\end{table}

\begin{table}
\caption{Goodness of the obtained fits to volume density pdfs at three evolutionary stages of simulated MC evolution and line-of-sight angle $\theta=84^\circ$. The notation is the same like in Table \ref{table_fit_goodness_obs}.} 
\label{table_fit_goodness_sim_rho} 
\begin{center}
\begin{tabular}{ccccrc@{~}c}
\hline 
\hline 
$t$ [ Myr ]& $N_{\rm bin}$ & $m$ & DF & $\chi^2$ & $\chi_{\rm crit,\,low}^2$ &  $\chi_{\rm crit,\,up}^2$ \\ 
\hline 
16.24 & 45 & 11	& 12	& ~2.93 & ~5.22 & 21.03\\	
18.59 & 42 & 10	& 12	& {\bf ~6.57} & ~5.23 & 21.03\\	
20.19 & 42 & ~7	& 21	& ~{\bf 11.84} & 11.59 & 32.67\\	
\hline 
\hline 
\end{tabular} 
\end{center}
\smallskip 
\end{table}

% \newpage

% \newpage

\newpage
\section{Derivation of density scaling law through a weighted fit}
\label{Appendix_fitting}

What should be an appropriate and physically reliable measure for the statistical weight $w_i$ of each scale $L_i$? Referring the reader to the suggested procedure (Sect. \ref{Fitting_procedure}), we point out that larger components have major contributions to the pdf fit while small components lead to minor local improvements. These contributions are proportional to the parameter $a_i$ and must be taken into account as one derives the scaling law of $\langle\rho\rangle$. Hence, the chosen estimate of the statistical weight of a scale is:
\begin{equation}
\label{eq_components_weight}
w_i(L_i) \propto a_i/\sum_i a_i \propto \sigma_i/\sum_i \sigma_i~.  
\end{equation}

\begin{figure} 
\begin{center}
\includegraphics[width=83mm]{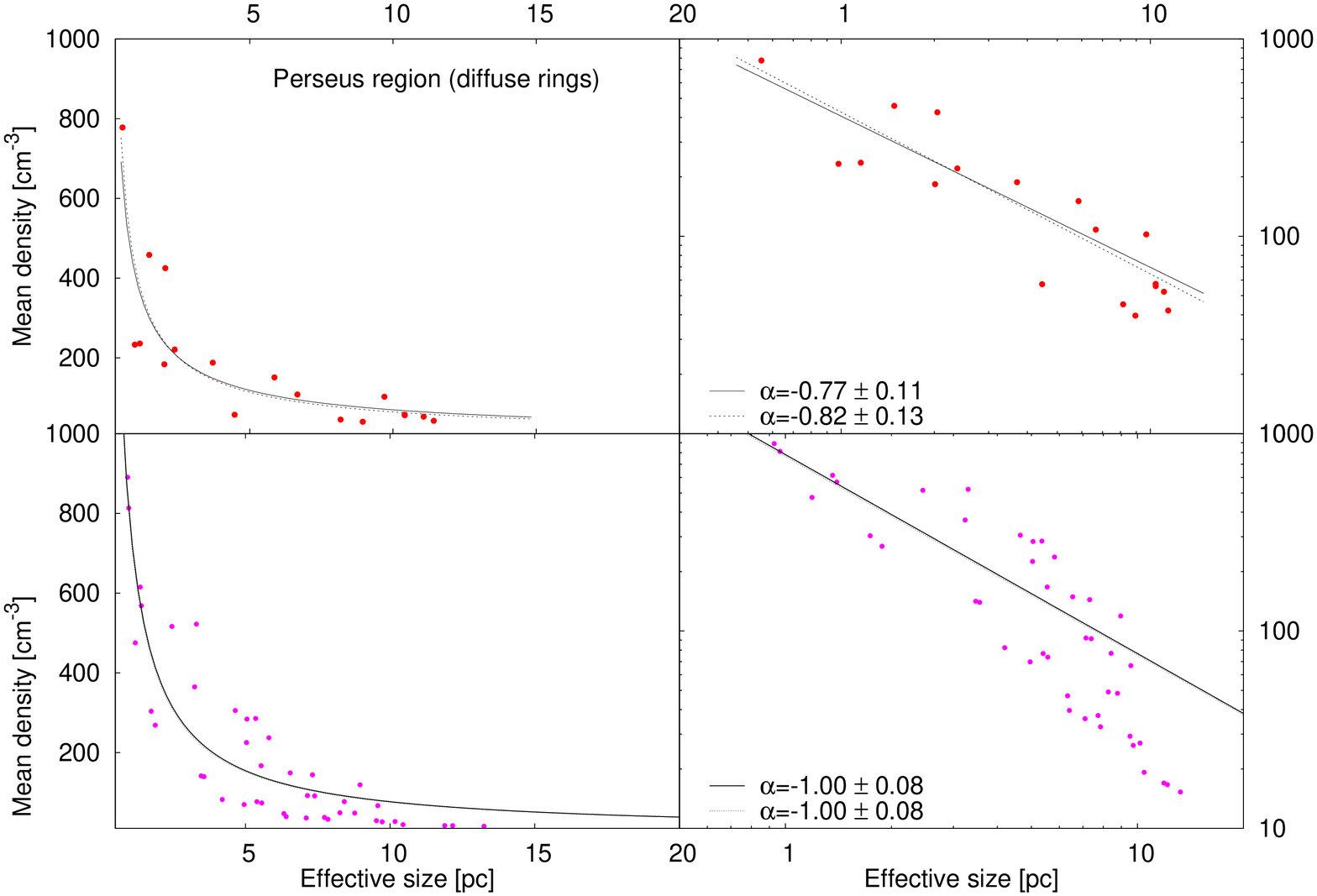}
\vspace{0.2cm}  
\caption{Correlations between effective size and mean density in linear (left) and log-log plot (right). Fits derived by standard NLLS procedure (dotted line) and by adopted weighting (solid line) are shown.}
\label{fig_n_scaling_linear_panel}
\end{center}
\end{figure}

Examples of fits derived in the diffuse rings of the Perseus region and in the simulated cloud are shown in Fig. \ref{fig_n_scaling_linear_panel}. Evidently, the adopted weighting yields similar slopes to those calculated from standard NLLS procedure.

\end{document}